\documentclass[aps,showpacs,nofootinbib,superscriptaddress,floatfix]{revtex4} 
\usepackage{amsmath}
\usepackage{graphicx,epsfig,wrapfig}
\usepackage{amsthm}
\usepackage{makeidx}
\usepackage{amsfonts}
\usepackage{amssymb}
\usepackage{natbib}
\usepackage{dsfont}
\usepackage{mathrsfs}
\usepackage{slashed}
\usepackage{slashbox}
\usepackage{color}

\definecolor{darkgreen}{rgb}{0,0.65,0}

\definecolor{green}{rgb}{0,.5,0}

\definecolor{orange}{rgb}{1,0.5,0}

\usepackage{epstopdf}
\usepackage{bm}
\usepackage[normalem]{ulem}
\usepackage{simplewick}
\renewcommand\sout{\bgroup \color[rgb]{0.55,0.00,0.99} \ULdepth=-.5ex \ULset}

\begin{document}

\newcommand{\mb}[1]{\mathbf{#1}}
\newcommand{\mc}[1]{\mathcal{#1}}
\newcommand{\non}{\nonumber}
\newcommand{\pep}[1]{\mathbf{#1}_{\perp}}
\newcommand{\pepgr}[1]{\bm{#1}_{\perp}}
\newcommand{\gdir}[1]{\gamma^{#1}}
\newcommand{\xk}{(x,\mathbf{k}_{\perp})}
\newcommand{\xkq}{(x,\mathbf{k}_{\perp}^{2})}
\newcommand{\pe}{(\mb{p}_{e})}
\newcommand{\xks}{(x,\pep{k};S)}
\newcommand{\pv}{\mathrm{P.V.}}
\newcommand{\xkl}{(x,\pep{k};\Lambda)}

\title{Revisiting the equivalence of light-front and covariant QED in the light-cone gauge}

\author{Luca Mantovani}
\email{luca.mantovani@pv.infn.it}
\affiliation{Dipartimento di Fisica, Universit\`a degli Studi di Pavia, I-27100 Pavia, Italy}
\affiliation{Istituto Nazionale di Fisica Nucleare, Sezione di
  Pavia,  I-27100 Pavia, Italy}

\author{Barbara Pasquini}
\email{barbara.pasquini@unipv.it}
\affiliation{Dipartimento di Fisica, Universit\`a degli Studi di Pavia, I-27100 Pavia, Italy}
\affiliation{Istituto Nazionale di Fisica Nucleare, Sezione di
  Pavia,  I-27100 Pavia, Italy}

\author{Xiaonu Xiong}
\email{Xiaonu.Xiong@pv.infn.it}
\affiliation{Istituto Nazionale di Fisica Nucleare, Sezione di
  Pavia,  I-27100 Pavia, Italy}

\author{Alessandro Bacchetta}
\email{alessandro.bacchetta@unipv.it}

\affiliation{Dipartimento di Fisica, Universit\`a degli Studi di Pavia, I-27100 Pavia, Italy}
\affiliation{Istituto Nazionale di Fisica Nucleare, Sezione di
  Pavia,  I-27100 Pavia, Italy}

\date{\today}


\allowdisplaybreaks[2]

\begin{abstract}
We discuss the equivalence between light-front time-ordered-perturbation theory and covariant quantum field theory in   
light-front quantization, in the case of quantum electrodynamics at one-loop
level. In particular, we review  the one-loop calculation of the vertex
correction, fermion self-energy and vacuum polarization. 
We apply the procedure of integration by residue over the light-front energy in
the loop to show how the perturbative expansion in covariant terms can be
reduced  to a sum of propagating and instantaneous diagrams of light-front
time-ordered perturbation theory. 
The detailed proof of equivalence between the two formulations of the theory
resolves the controversial question on which form should be used for the
gauge-field propagator in the light-cone gauge in the covariant approach.  

\end{abstract}

\maketitle
The light-front formulation of quantum field theory has remarkable advantages
with respect to the canonical equal-time (instant-form) approach (see, e.g.,
\cite{Brodsky:1997de}).  
The equivalence between the two approaches has been the subject of several
works~\cite{Weinberg:1966jm,Chang:1968bh,Kogut:1969xa,Chang:1973qi,Brodsky:1973kb,Ligterink:1994tm,Bakker:2005mb}, 
with applications 
 to scalar $\phi^3$ theory \cite{Sawicki:1991sr}, Yukawa
 theory~\cite{Schoonderwoerd:1998iy,Schoonderwoerd:1998qj}, scalar Quantum
 ElectroDynamics (QED) \cite{Ji:2014vha} and standard QED
 \cite{Misra:2005dt,Patel:2010vd}.  

In this work, we will focus on the proof of the equivalence in QED at one-loop
level. In particular, we will present the calculation of the one-loop diagrams
in QED in two different approaches: 
 1) using light-front time-ordered perturbation theory (TOPT), where the
 contribution from each light-front time-ordered graph is calculated according
 to the rules of old-fashioned Hamiltonian perturbation
 theory~\cite{Kogut:1969xa,Lepage:1980fj}; 
 2) using the covariant formulation of QED in light-front quantization.

The main difficulty in the correspondence between the two formulations of the
theory 
 is  to match the presence of instantaneously propagating particles.
This issue becomes particularly subtle when dealing with gauge theories. In
light-front quantization, one commonly works in the non-covariant
light-cone gauge $n \cdot A=0$, with $n$ being an
external light-like vector 
$n^2=0$ \cite{Srivastava:2000cf}. 
With this gauge choice, 
it becomes nontrivial to determine the detailed form of the gauge field
propagator 
(i.e., the 
photon propagator in QED) and to match it with time-ordered diagrams  
containing instantaneous photons
\cite{Misra:2005dt,Ji:2014vha}.

When dealing with the covariant approach, in the literature one encounters two
different expressions for the gauge field propagator $D^{\mu\nu}(q)$ in the
light-cone gauge. The first one, containing the sum of two terms, reads
\cite{Kogut:1969xa,Harindranath:1996hq}: 
\begin{equation}
 \mc{D}^{\mu\nu}(q)=\frac{-i}{q^2}\left(g^{\mu\nu}-\frac{q^\mu n^\nu+q^\nu
     n^\mu}{q^+}\right) \; . \label{twoterm}\end{equation} 
This expression can be recovered by adding to the Lagrangian density for
the free vector gauge field a gauge-fixing term of the form
$\left(n_{\mu}A^{\mu}\right)^2$, as shown in
Ref.~\cite{Suzuki:2003fm}; but this is not enough to fix the gauge completely,
and as a result both the longitudinal and the transverse photon degrees of
freedom propagate.\\ 
Another expression for the gauge-field propagator is often used, containing the sum of three terms \cite{Mustaki:1990im}:
\begin{equation} 
\mc{D}^{\mu\nu}_{T}(q)=\frac{-i}{q^2}\left(g^{\mu\nu}-\frac{q^\mu n^\nu+q^\nu
    n^\mu}{q^+} +q^2\frac{n^{\mu}n^{\nu}}{\left(q^{+}\right)^{2}} \right)\;
. \label{threeterm}\end{equation} 
This expression can in turn be recovered by adding to the Lagrangian density a
gauge-fixing term of the form $\left(n_{\mu}A^{\mu}\right)^2 +
\left(\partial_{\mu}A^{\mu}\right)^2$ \cite{Suzuki:2003fm}, which completely
fixes the gauge; this is equivalent to considering only the transverse degrees
of freedom as the propagating ones. 
However, the use of a non-covariant axial gauge, such as the light-cone gauge,
generates a new contribution to the interaction Hamiltonian, corresponding to
an instantaneous interaction. The last term of Eq.~\eqref{threeterm},
which comes from removing the longitudinal degree of freedom,
compensates for this extra-term describing the instantaneous interaction from
the Hamiltonian \cite{Misra:2005dt,Ji:2014vha}. Therefore, it is possible to
use either the three term expression of Eq.~\eqref{threeterm} together with
the instantaneous interaction in the Hamiltonian, or
the two-term expression of
Eq.~\eqref{twoterm}, omitting the instantaneous interaction in the
Hamiltonian.  

This ambiguity has been a source of confusion in establishing which
form should
be used, and is a relevant issue when
proving the equivalence between the covariant perturbation theory and
light-front TOPT, in particular when matching with the contribution
from instantaneous photons. 

We therefore believe that it is illustrative to re-derive the proof of the
equivalence between covariant and light-front TOPT approach at one-loop level
in QED, trying to clarify a few misleading results in literature. 
Our findings are partially at variance with previous calculations~\cite{Misra:2005dt,Patel:2010vd,Ji:2014vha}, and we will  point out 
 where the origin  of differences is. 
\\
The  paper is organized as follows: in Sec.~\ref{basics} we discuss the
structure of the QED Hamiltonian, both in light-front quantization
\cite{Mustaki:1990im} and in covariant theory in the light-cone gauge. In
Sec.~\ref{triangle} we discuss the QED vertex correction, first in the
light-front TOPT and then in the covariant approach with light-front
coordinates, showing the equivalence between the two formulations. In
Sec.~\ref{selfen} we apply the same scheme to the self-energy diagrams of
both the electron and the photon.  
In particular, we
show how the one-loop expressions for the vertex correction, fermion
self-energy, and vacuum polarization in 
the covariant perturbation theory can be reduced to a sum of propagating and
instantaneous diagrams of 
light-front time-ordered perturbation theory by using the well-known technique
of  integration by residue over the light-front energy in the loop. 
Finally in Sec.~\ref{conclusions} we summarize our results and draw our
conclusions. Technical details are collected in the appendices.

\section{QED Hamiltonian in the light-front}\label{basics}

Our starting point is the gauge- and Lorentz-invariant QED Lagrangian density:
\begin{equation}
\mathscr{L}_{QED}=\bar{\psi}(i\slashed{\partial}-m)\psi-\frac{1}{4}F^{\mu\nu}F_{\mu\nu}+e\bar \psi\gamma^\mu\psi A_{\mu} \; , \label{qedlagr}
\end{equation}
where $e >0$ and $m$ are the electron charge and  mass, respectively, and $F^{\mu\nu}=\partial^{\mu}A^{\nu}-\partial^{\nu}A^{\mu}$ is the photon field tensor.

We first describe the approach   in light-front quantization, 
revisiting the results of Refs.~\cite{Mustaki:1990im,Srivastava:2000cf}. We introduce light-front coordinates $a^{\mu}=\left(a^{+},a^{-},\pep{a}\right)$ for a generic four-vector $a^{\mu}$, defining $a^{+}=\left(a^{0}+a^{3}\right)/\sqrt{2}$, $a^{-}=\left(a^{0}-a^{3}\right)/\sqrt{2}$, and $\pep{a}=(a^1,a^2)$. By using the equations of motion in light-front coordinates and working in the light-cone gauge $n_{\mu}A^{\mu}=0$, with $n^{\mu}=\left(0,1,\pep{0}\right)$, the light-front QED Hamiltonian 
can be written as 
\begin{equation}
H=\int d^{2}\pep{x}dx^{-}\,T^{+-}=H_{0}+V_{1}+V_{2}+V_{3},\label{hamiltonian}
\end{equation}
where $H_{0}$ is the free Hamiltonian and
the interaction terms are given by
\begin{subequations}
\begin{align} 
V_{1}=&e\int d^{2}\pep{x}dx^{-}\,\bar{\xi}(x)\gamma^{\mu}\xi(x) a_{\mu}(x) \; , \label{V1} \\
V_{2}=&-\frac{ie^{2}}{4}\int d^{2}\pep{x}dx^{-}dy^{-}\,\Theta\left(x^{-}-y^{-}\right)\bar{\xi}(x)\gamma_{i}a^{i}(x)\gamma^{+}\gamma_{j}a^{j}(y)\xi(y) \; , \label{V2}\\
V_{3}=&-\frac{e^{2}}{4}\int d^{2}\pep{x}dx^{-}dy^{-}\,\bar{\xi}(x)\gamma^{+}\xi(x)\left| x^- -y^-\right|\bar{\xi}(y)\gamma^{+}\xi(y) \; , \label{V3}
\end{align}
\end{subequations}
with $y=\left(x^{+},y^-,\pep{x}\right)$. Eqs.~\eqref{V1}-\eqref{V3} are written in terms of the independent degrees of freedom in light-front quantization: the photon field $a^\mu=(0,a^-,\pep{A})$ (with $\partial_{+}a^{-}=\pepgr{\partial}\cdot\pep{A}$) is rewritten in terms of the transverse components $A^{i}$, while $\xi(x)=\Lambda_{+}\psi(x)$ (with $\Lambda_+ =\frac{1}{2}\gamma^{-}\gamma^+$) are the ``good'' components of the fermion field. The interaction term $V_1$ in  Eq.~\eqref{V1} describes a standard QED three-point vertex; 
$V_2$ in Eq.~\eqref{V2} and $V_3$ in Eq.~\eqref{V3}, instead, are the non-local four-point vertices, corresponding to the exchange of an instantaneous fermion and photon, respectively. From the Hamiltonian \eqref{hamiltonian}, one can derive the light-front time-ordered contributions to one-loop processes in QED, as we will show in the following sections.

As we aim to match the description in light-front TOPT with the covariant theory of QED, we need to consider how the usual Feynman rules for the gauge-field propagator should be modified in the non-covariant light-cone gauge. 

The photon propagator in the light-cone gauge $A^+=0$ can be derived from the free theory on the light-front, where the additional Lorenz condition $\partial_{\mu}A^{\mu}=0$ is obtained from the equation of motion \cite{Srivastava:2000cf,Ji:2014vha}. The constraints $A^{+}=0$ and $\partial_{\mu}A^{\mu}=0$ restrict the dynamical degrees of freedom for the polarization vectors of the photon only to the transverse ones.
This leads to the following form for the gauge-field propagator~\cite{Srivastava:2000cf}
\begin{equation} 
\mc{D}_T^{\mu\nu}(q)=\frac{-i}{q^2}\left(g^{\mu\nu}-\frac{q^\mu n^\nu+q^\nu n^\mu}{q^+}+q^2\frac{n_{\mu}n_{\nu}}{\left(q^{+}\right)^{2}}\right) \; , \label{proplc}\end{equation}
which is proportional to the sum over the transverse-polarization of the photon degrees of freedom
\begin{equation} d_{T}^{\mu\nu}(q)=\sum_{\lambda=1}^2 \varepsilon^{\mu\,*}_{\lambda}(q)\varepsilon^{\nu}_{\lambda}(q)=-g^{\mu\nu}+\frac{q^\mu n^\nu+q^\nu n^\mu}{q^+}-q^2\frac{n^{\mu}n^{\nu}}{\left(q^{+}\right)^{2}} .  \label{dT}
\end{equation}
The same result is obtained in Ref.~\cite{Suzuki:2003fm} by adding to the free photon Lagrangian the gauge 
fixing terms 
\begin{equation} 
-\frac{1}{2\alpha}\left(n_{\mu}A^{\mu}\right)^2 -\frac{1}{2\beta} \left(\partial_{\mu}A^{\mu}\right)^2 \; , 
\label{gauge-fixing}
\end{equation}
corresponding, in order,  to the gauge-fixing condition for the light-cone gauge and for the Lorenz gauge. \\
Notice that the manifest Lorentz invariance is spoiled 
by the use of a non-covariant axial gauge; indeed, the photon propagator \eqref{proplc} itself has two non-covariant extra terms, in addition to the $-g^{\mu\nu}$ term of the manifestly-covariant Feynman gauge. 

In the interacting theory the equation of motion for the gauge field is
\begin{equation}
(\partial^{+})^2 A^-=\partial^+ \partial^{i}A_i -e\bar{\psi}\gamma^{+}\psi \, \label{eomint}
\end{equation}
It does not lead to the Lorenz condition $\partial_{\mu}A^{\mu}=0$, but still reduces the independent degrees of freedom of the photon to the transverse ones.

Using Eq.~\eqref{eomint} to eliminate $A^{-}$ in the interaction Hamiltonian, we obtain
\begin{eqnarray}
H=H_{0} + V+ V_I \; . \label{h-cov}
\end{eqnarray}
Here $V$ is the three-point vertex
\begin{equation} V = e\int d^{2}\pep{x}dx^{-}\,\bar{\psi}(x)\gamma^{\mu}\psi(x) a_{\mu}(x) \; , \label{V} 
\end{equation}
where $a^{\mu}$ is the same used in Eqs.~\eqref{V1}-\eqref{V3}, while $V_I$ is given by
\begin{equation}
V_{I}=-\frac{e^{2}}{4}\int d^{2}\pep{x}dx^{-}dy^{-}\,\bar{\psi}(x)\gamma^{+}\psi(x)\left| x^- -y^-\right|\bar{\psi}(y)\gamma^{+}\psi(y) \; . \label{Vi}
\end{equation}
We will refer to $V_{I}$ as the ``instantaneous interaction'' term. It can be shown that, in physical processes, the contribution from the  third term of the propagator in Eq.~\eqref{proplc} is exactly compensated for by the contribution from the instantaneous interaction term in the Hamiltonian (see e.g. in Ref.~\cite{Srivastava:2000cf}). 
Notice that in Eqs.~\eqref{V} and \eqref{Vi} we did not rewrite the fermion fields in terms of the dynamical fields in light-front quantization, as we did in Eqs.~\eqref{V1}-\eqref{V3}. Indeed, the presence of the instantaneous interaction occurs in covariant quantization for any generic axial gauge $n_{\mu}A^{\mu}=0$, as shown in Ref.~\cite{Ji:2014vha} for the scalar QED case (which can be easily extended to the QED case as well). 
This situation is well known also in the case of the Coulomb gauge \cite{Weinberg:1995mt} and does not therefore depend on the use of light-front rather than instant-form coordinates.
For this reason, it is natural to consider as the starting point for the covariant calculation the three-term propagator \eqref{proplc}, including also the diagram containing the instantaneous interaction. Alternatively, we can drop the third term and consider only the ordinary three-point vertex \eqref{V},
neglecting the instantaneous interaction at the same time.

\section{One-loop QED vertex correction}\label{triangle}

In this section we discuss the vertex correction at one-loop level in QED. We first  revisit the calculation in light-front TOPT~\cite{Weinberg:1966jm,Brodsky:1973kb,Mustaki:1990im}, then we present the results in covariant theory in light-front coordinates. We show
how to derive light-front time-ordered 
diagrams from a given Feynman diagram by integrating
over the light-front energy.
In showing the equivalence between the different formulations of the theory, we will discuss the form of the photon propagator.

\subsection{Light-front Time-ordered perturbation theory}

The vertex correction in light-front TOPT originates from contributions of the interaction terms in the form  $(V_{1})^3$, $V_{1}V_{2}$ and $V_{1}V_{3}$, see Eqs.~\eqref{V1}-\eqref{V3}.

\begin{figure}[h]
\includegraphics[width=12cm]{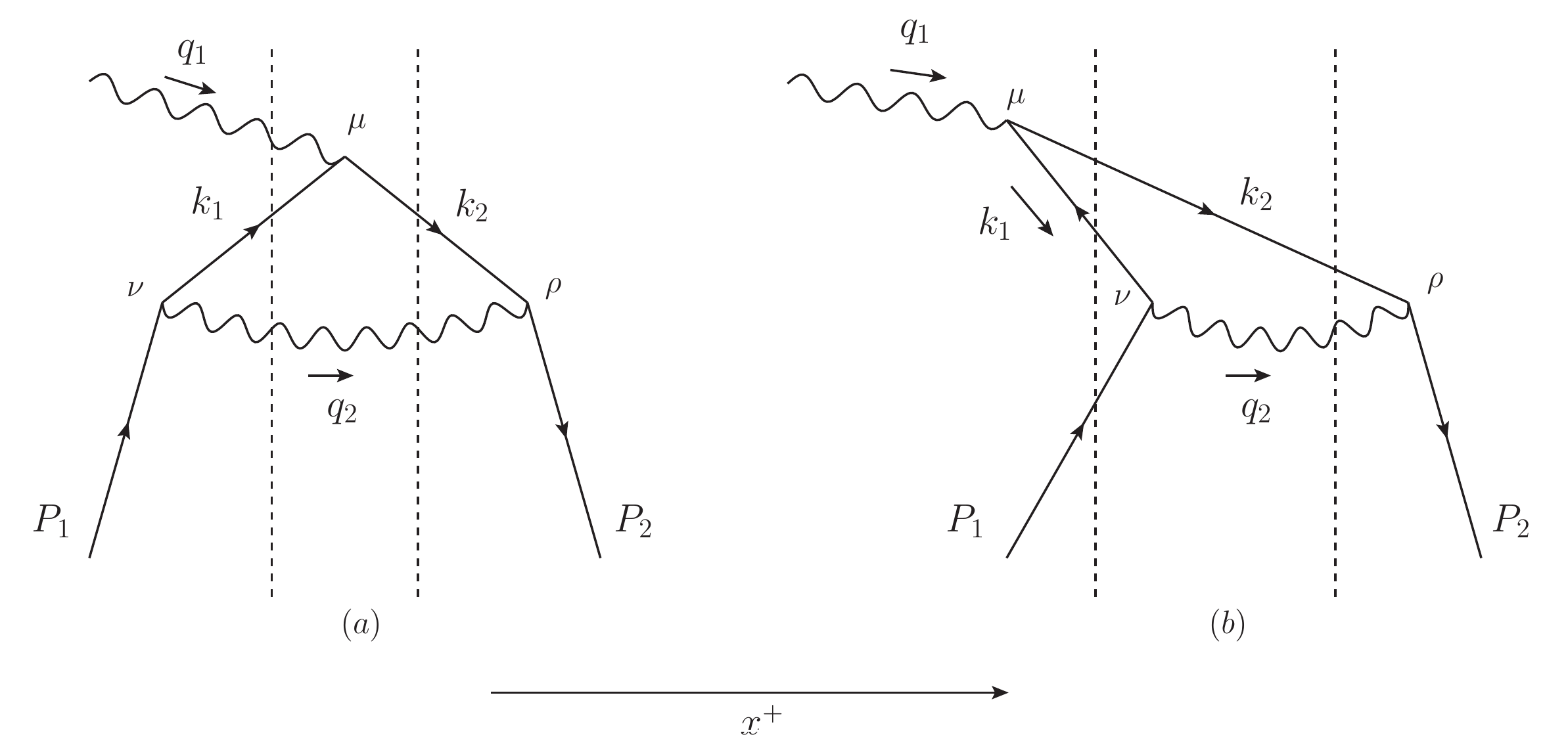}
\caption{Diagrams for the vertex correction in light-front TOPT at one-loop order.
The vertical dashed lines are at fixed light-front time.}
\label{lctopt}
\end{figure}

\begin{figure}[h]
\includegraphics[width=13cm]{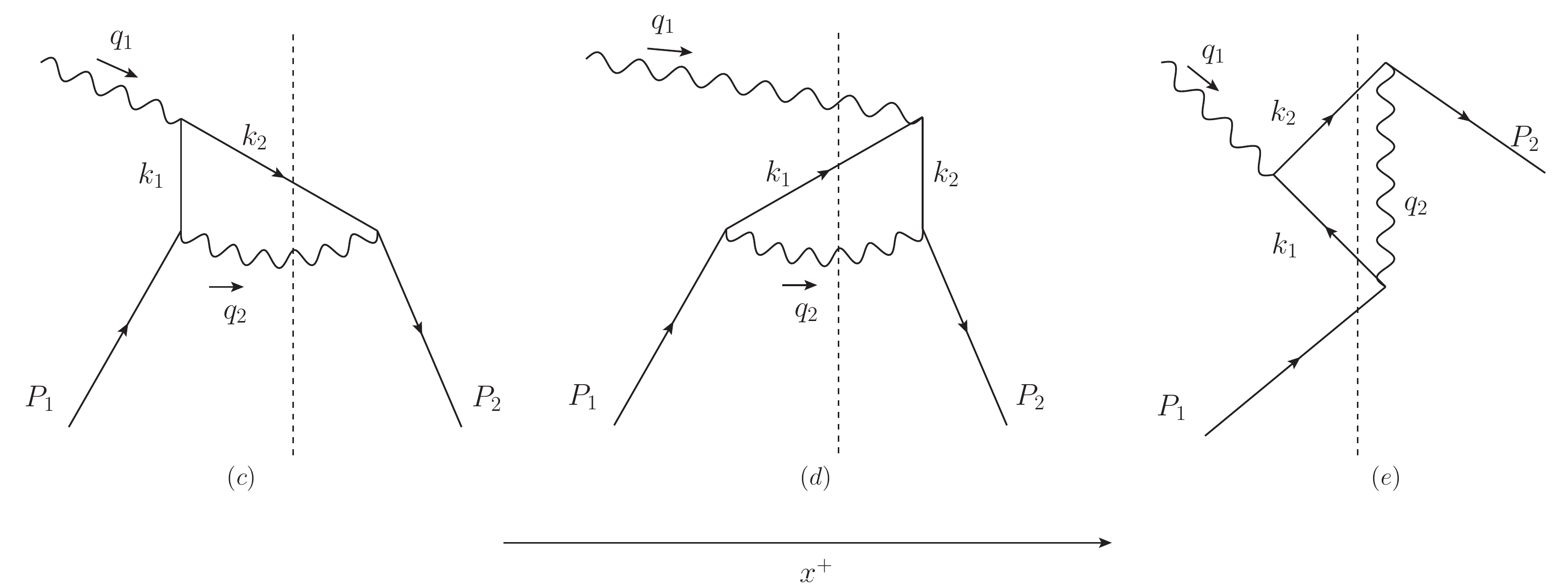}
\caption{Diagrams for the vertex correction in light-front TOPT at one-loop order
with instantaneous fermion exchange ((c) and (d)) and instantaneous photon exchange ((e)).
The vertical dashed lines are at fixed light-front time.}
\label{instdiag}
\end{figure}

The $(V_{1})^3$ contribution brings the two diagrams shown in Fig.~\ref{lctopt}, which are distinct for different ordering of the interaction vertexes with light-front time; note that in diagram (b) there appears a positron, labeled with an arrow of opposite direction with respect to its plus momentum. All other possible time-ordered diagrams, with disconnected contributions of pair production or annihilation from the vacuum
vanish in the infinite-momentum frame;
nonetheless, the presence of the $V_{2}$ and $V_{3}$ interaction terms in the Hamiltonian leads to additional diagrams, containing instantaneous fermions and photon propagators. This means that we actually need to consider five time-ordered diagrams including the three instantaneous interaction diagrams shown in Fig.~\ref{instdiag}. It should be noticed that diagram (c) of Fig.~\ref{instdiag} is peculiar, since it can be recovered from both diagram (a) and (b) of Fig.~\ref{lctopt}, by considering the propagator with momentum $k_1$ as instantaneous. Diagrams (d) and (e), on the contrary, can be drawn only as limits of diagrams (a) and (b), respectively. 

In the calculation of the scattering amplitude, 
we set the momenta of the particles as shown in the corresponding diagrams. We recall that, in the TOPT approach, all particles are on-shell; however, although overall conservation of four-momentum between initial and final state still holds, energy (i.e. the minus component of momenta in light-front form) is not conserved in each interaction vertex.
As we aim to match the terms derived from the evaluation of the time-ordered diagrams with the corresponding current obtained in the covariant approach, it is convenient to decompose the momentum $k^\mu$ of a particle of mass $m$ into an on-shell and an off-shell component, by defining
\begin{subequations}
\begin{align}
& k^\mu_{\mathrm{on-shell}}\equiv \tilde{k}^\mu = \left(k^{+},\frac{\pep{k}^{2}+m^{2}}{2k^{+}},\pep{k}\right), \label{splitting1} \\
& k_{\mathrm{off-shell}}^{\mu}\equiv \hat{k}^\mu= \left(0,\frac{k^{2}-m^{2}}{2k^{+}},\pep{0}\right)=\frac{k^{2}-m^{2}}{2k^{+}}n^{\mu} \; , \label{splitting2}
\end{align}
\end{subequations}
such that $k^{\mu}=\tilde{k}^\mu+\hat{k}^\mu$.

For convenience, we switch to the symmetric notation for the momenta, by setting:
\begin{equation} q_1\equiv\Delta\; , \quad P_{1,2}=P\mp\frac{\Delta}{2}\; , \quad k_{1,2}=k\mp\frac{\Delta}{2} \; ,\quad q_2 = P-k \; .\label{symmnot}
\end{equation}
The last two identities in the above expression are consequence of momentum conservation and it is hence understood that they are valid only for the plus and transverse (but not for the minus) components when working within  light-front TOPT.
We also define the longitudinal momentum fraction $x:\,=k^+/P^+$ and the skewedness parameter $\xi:\,=\Delta^+/2P^+$. Furthermore, without loss of generality, we can choose the reference frame where $\pep{P}=\pep{0}$ and fix $\xi>0$, which means we are considering an incoming photon (i.e. $\Delta^+>0$).
The on-shell conditions give:
\begin{equation} P^-=\frac{1}{2P^+}\left(m^2-\frac{t}{4}\right)\; , \quad \Delta^-=-2\xi P^- \; , \quad k_{1,2}^{-}=\frac{\mb{k}_{1,2_{\perp}}^{2}+m^{2}}{2k_{1,2}^{+}}\; , \quad q_{2}^{-}=\frac{\mb{q}^2_{2_{\perp}}}{2q_{2}^{+}} \; , \label{onshell}
\end{equation} with $t\equiv q^2$ and
\begin{equation} k_{1,2}^{+}=k^{+}\mp\frac{\Delta^{+}}{2}\; , \quad \mb{k}_{1,2_{\perp}}= \pep{k}\mp\frac{\pep{\Delta}}{2} \; , \quad q_{2}^{+}=P^{+}-k^{+}\; , \quad \mb{q}_{2_{\perp}}=-\pep{k}\; .\label{ptcons} \end{equation}

Let us start from the contribution of order $V_1^3$ to the scattering amplitude; the corresponding matrix element of the transition matrix $iT$ is:
\begin{align}
\Lambda^{TOPT}_{(V_1^3)}=\langle e_{P_{2}}\lvert iT \rvert e_{P_1}\gamma_{q_1}\rangle = i \left\langle e_{P_{2}}\left\lvert V_{1}\frac{1}{P^-_i -H_0}V_1\frac{1}{P^-_i -H_0} V_1\right\rvert e_{P_1}\gamma_{q_1}\right\rangle \label{lambdatopt}
\end{align}
where $P^-_i=P_{1}^{-}+q_1^-$ is the total energy of the initial states. Spin indices are omitted as they are not relevant in the current discussion. By inserting the resolution of the identity with the complete set of eigenstates of the free Hamiltonian, it is possible to identify the contributions $\Lambda^{TOPT}_{(a)}$ and $\Lambda^{TOPT}_{(b)}$ corresponding to diagrams (a) and (b) of Fig.~\ref{instdiag}, i.e. 
  \begin{equation}
\Lambda^{TOPT}_{(V_1^3)}=\Lambda^{TOPT}_{(a)}+\Lambda^{TOPT}_{(b)}.
\end{equation}
The first contribution can be recast in the form
\begin{equation}
\Lambda_{(a)}^{TOPT}=(2\pi)^3\delta\left(P_1^+ +q_1^+ -P_2^+\right)\delta^{(2)}\left(\pep{P_1}+\pep{q_1}-\pep{P_2}\right)\left(-ie\right)\varepsilon_{\mu}(q_1){J}_{(a)}^\mu \; . \label{jdefa} 
\end{equation}

For later convenience , we isolate the energy denominator of the current $J^{\mu}_{(a)}$:
\begin{align} D_{(a)} = & \left\langle \gamma_{q_{1}}e_{k_{1}}\gamma_{q_{2}}\left|\frac{1}{P_{1}^{-}+q_1^- -H_0^{-}}\right|\gamma_{q_{1}}e_{k_{1}}\gamma_{q_{2}}\right\rangle \left\langle e_{k_{2}}\gamma_{q_{2}}\left|\frac{1}{P_{1}^{-}+q_1^- -H_0^{-}}\right|e_{k_{2}}\gamma_{q_{2}}\right\rangle \non \\
= & \frac{1}{P_{1}^{-}+q_{1}^{-}-\left(q_{1}^{-}+k_{1}^{-}+q_{2}^{-}\right)}\frac{1}{P_{1}^{-}+q_{1}^{-}-\left(k_{2}^{-}+q_{2}^{-}\right)} \; . \label{Da}\end{align}
By switching to the symmetric notation and using Eqs.~\eqref{onshell}-\eqref{ptcons}, one can rewrite:
\begin{align}
P_{1}^{-}+q_{1}^{-}-\left(q_{1}^{-}+k_{1}^{-}+q_{2}^{-}\right)&=-\frac{(1-x)\,\pep{k}\cdot\pep{\Delta}-(1-\xi)\,\pep{k}^2-(1+\xi)(1-x)^2\left(m^2-\frac{t}{4}\right)}{2P^+(1-x)(x-\xi)}
=\kappa_3-\kappa_1 \; , \label{k3k1}\\
P_{1}^{-}+q_{1}^{-}-\left(k_{2}^{-}+q_{2}^{-}\right)&
=-\frac{(1-x)\,\pep{k}\cdot\pep{\Delta}+(1+\xi)\,\pep{k}^2+(1-\xi)(1-x)^2\left(m^2-\frac{t}{4}\right)}{2P^+(1-x)(x+\xi)}=\kappa_{3}-\kappa_{2} \; , \label{k3k2}\end{align}
where we have defined
\begin{subequations}
\begin{align} & \kappa_{1,2}^-=\frac{\pep{k}^2\mp\pep{k}\cdot\pepgr{\Delta}+(1\mp\xi x)\left(m^2-\frac{t}{4}\right)}{2P^+(x\mp\xi)} \; , \label{poles1} \\
& \kappa_{3}^- = \frac{\pep{k}^2+(x-1)\left(m^2-\frac{t}{4}\right)}{2(x-1)P^+}. \label{poles2}\end{align}
\end{subequations}
The numerator of the current $J^\mu_{(a)}$ is given by
\begin{equation}
N_{(a)} = -\frac{e^{2}P^+}{\left(2\pi\right)^{3}}\int_{\xi}^{1} dx\int d^{2}\boldsymbol{k}_{\perp}\frac{1}{\left(2P^+\right)^3(1-x)\left(x^{2}-\xi^2\right)}\bar{u}(P_2)\gamma^{\rho}\left(\tilde{k}\!\!\!/_{2}+m\right)\gamma^{\mu}\left(\tilde{k}\!\!\!/_{1}+m\right)\gamma^{\nu}d_{\nu\rho}\left(\tilde{q}_{2}\right)u(P_1) \; . \label{numa}\end{equation}
where $\tilde k\!\!\!/+m=u(\tilde k)\bar{u}(\tilde k)$, and the limits of integration in the variable $x$ are given by the constraint that all particles move along the positive light-front direction.\\
 In Eq.~\eqref{numa},  $d^{\mu\nu}(q)$ is given by the transverse-polarization sum of the intermediate photon taken on shell, i.e.
\begin{equation} d^{\mu\nu}(q)=-g^{\mu\nu}+\frac{q^{\mu}n^{\nu}+q^{\nu}n^{\mu}}{q^{+}} \;. \label{d2}\end{equation}
This result is a direct consequence of taking the intermediate photon on-shell ($q^2=0$), according to the rules of old-fashioned perturbation theory, and is at variance with the expression used in the calculation of light-front time-ordered diagrams of Ref.~\cite{Ji:2014vha} which uses the three-term sum of Eq.~\eqref{dT}.
\\
Collecting the results from Eqs.~\eqref{numa} and \eqref{Da}, we finally obtain the following contribution of 
the diagram (a) to the current
\begin{equation}
J^\mu_{(a)} = -\frac{e^{2}P^+}{\left(2\pi\right)^{3}}\int_{\xi}^{1} dx\int d^{2}\boldsymbol{k}_{\perp}\frac{1}{\left(2P^+\right)^3(1-x)\left(x^{2}-\xi^2\right)}\bar{u}(P_2)\gamma^{\rho}\frac{\left(\tilde{k}\!\!\!/_{2}+m\right)}{\kappa_3-\kappa_1}\gamma^{\mu}\frac{\left(\tilde{k}\!\!\!/_{1}+m\right)}{\kappa_{3}-\kappa_{2}}\gamma^{\nu}d_{\nu\rho}\left(\tilde{q}_{2}\right)u(P_1) \; .
\label{ja}
\end{equation}

A similar procedure can be followed also for diagram (b) in Fig.~\ref{lctopt}, corresponding to the contribution
\begin{equation}
\Lambda_{(b)}^{TOPT}=(2\pi)^3\delta\left(P_1^+ +q_1^+ -P_2^+\right)\delta^{(2)}\left(\pep{P_1}+\pep{q_1}-\pep{P_2}\right)\left(-ie\right)\varepsilon_{\mu}(q_1){J}_{(b)}^\mu \; . \label{jdefb} 
\end{equation}
The  current $J_{(b)}^{\mu}$ has an energy denominator given by
\begin{align} D_{(b)} = & \left\langle e_{P_{1}}e^+_{k_{1}}e_{k_{2}}\left|\frac{1}{P_{1}^{-}+q_1^- -H^{-}}\right|e_{P_{1}}e^+_{k_{1}}e_{k_{2}}\right\rangle \left\langle e_{k_{2}}\gamma_{q_{2}}\left|\frac{1}{P_{1}^{-}+q_1^- -H^{-}}\right|e_{k_{2}}\gamma_{q_{2}}\right\rangle \non \\
= & \frac{1}{P_{1}^{-}+q_{1}^{-}-\left(P_{1}^{-}-k_{1}^{-}+k_{2}^{-}\right)}\frac{1}{P_{1}^{-}+q_{1}^{-}-\left(k_{2}^{-}+q_{2}^{-}\right)} \; .\end{align}
We can rewrite:
\begin{equation}
P_{1}^{-}+q_{1}^{-}-\left(P_{1}^{-}-k_{1}^{-}+k_{2}^{-}\right) =\frac{x\,\pep{k}\cdot\pep{\Delta}-\xi\,\pep{k}^2-\xi(1-x^2)\left(m^{2}-\frac{t}{4}\right)}{(x^2-\xi^2)P^+} =\kappa_{2}-\kappa_{1} \; ,\label{k2k1}
\end{equation}
while the energy denominator for the second intermediate state is again given by Eq.~\eqref{k3k2}. 
If we evaluate also the numerator of $J_{(b)}^{\mu}$, we come up with the final result:
\begin{equation} J^\mu_{(b)} = -\frac{e^{2}P^+}{\left(2\pi\right)^{3}}\int_{-\xi}^{\xi} dx\int d^{2}\boldsymbol{k}_{\perp}\frac{1}{\left(2P^+\right)^3(1-x)\left(x^{2}-\xi^2\right)}\bar{u}(P_2)\gamma^{\rho}\frac{\left(\tilde{k}\!\!\!/_{2}+m\right)}{\kappa_{3}-\kappa_{2}}\gamma^{\mu}\frac{\left(\tilde{k}\!\!\!/_{1}+m\right)}{\kappa_{2}-\kappa_{1}}\gamma^{\nu}d_{\nu\rho}\left(\tilde{q}_{2}\right)u(P_1) \; . \label{jb}
\end{equation}
We now focus on the diagrams containing instantaneous propagators, which come from $V_1V_2$, $V_2V_1$
and $V_1V_3$ interaction terms in the  matrix element.

The result for the current $J^{\mu}_{(c)}$ of 
The diagram (c) in Fig.~\ref{instdiag} corresponds to the matrix element

\begin{align}
\Lambda^{TOPT}_{(V_1V_2)} &= i \left\langle e_{P_{2}}\left\lvert V_{1}\frac{1}{P^-_i -H_0} V_2\right\rvert e_{P_1}\gamma_{q_1}\right\rangle \\
&=(2\pi)^3\delta\left(P_1^+ +q_1^+ -P_2^+\right)\delta^{(2)}\left(\pep{P_1}+\pep{q_1}-\pep{P_2}\right)\left(-ie\right)\varepsilon_{\mu}(q_1){J}_{(c)}^\mu \; . \label{jcdef} 
\end{align}
Following a similar procedure as above, one finds that the current $J^{\mu}_{(c)}$ is given by
\begin{equation}
J^{\mu}_{(c)}= -\frac{e^2 P^+}{\left(2\pi\right)^3}\int_{-\xi}^1 dx\int d^2\pep{k}\,\frac{1}{\left(2P^+\right)^3(1-x)
\left(x^{2}-\xi^2\right)}\bar{u}(P_2)\gamma^{\rho}\frac{\left(\tilde{k}\!\!\!/_{2}+m\right)}{\kappa_{3}-\kappa_{2}}
\gamma^{\mu}\gamma^+\gamma^{\nu}d_{\nu\rho}\left(\tilde{q}_{2}\right)u(P_1) \; . \label{jc}
\end{equation}
Note that this result can also be obtained by adding together  the contributions from diagrams (a) and (b), replacing $(\slashed{k}_1+m)\mapsto \gamma^+$ and taking away from them the energy denominators \eqref{k3k1} and \eqref{k2k1}, respectively.
This is consistent with the TOPT rules for an instantaneous fermion propagator~\cite{Lepage:1980fj}.

Similarly, we obtain the following result for the contributions to the current from diagrams (d) in Fig.~\ref{instdiag}
\begin{equation}
J^{\mu}_{(d)}= -\frac{e^2 P^+}{\left(2\pi\right)^3}\int_{\xi}^1 dx\int d^2\pep{k}\,\frac{1}{\left(2P^+\right)^3(1-x)\left(x^{2}-\xi^2\right)}\bar{u}(P_2)\gamma^{\rho}\gamma^{+}\gamma^{\mu}\frac{\left(\tilde{\slashed{k}_{1}}+m\right)}{\kappa_{3}-\kappa_{1}}\gamma^{\nu}d_{\nu\rho}\left(\tilde{q}_{2}\right)u(P_1) \;.  \label{jd} 
\end{equation}

\noindent
Finally, the diagram (e) in Fig.~\ref{instdiag} with instantaneous photon can be obtained from Eq.~\eqref{jb} with the substitution $d^{\mu\nu}(q)/q^+\mapsto n^\mu n^\nu/(q^{+})^{2}$ and removing the term $\kappa_2-\kappa_1$ from the energy denominator \eqref{jb}. As a result, it reads

\begin{equation}
J^{\mu}_{(e)}= -\frac{e^2 P^+}{\left(2\pi\right)^3}\int_{-\xi}^{\xi} dx\int d^2\pep{k}\,\frac{1}{4\left(P^+\right)^3(1-x)\left(x^{2}-\xi^2\right)}\bar{u}(P_2)\gamma^{\rho}\frac{\left(\tilde{k}\!\!\!/_{2}+m\right)}{\kappa_{3}-\kappa_{1}}\gamma^{\mu}\left(\tilde{\slashed{k}_{1}}+m\right)\gamma^{\nu}\frac{n_{\nu}n_{\rho}}{(1-x)P^{+}}u(P_1) \; . \label{je}
\end{equation}

\subsection{Covariant approach}

\begin{figure}[h]
\includegraphics[width=10cm]{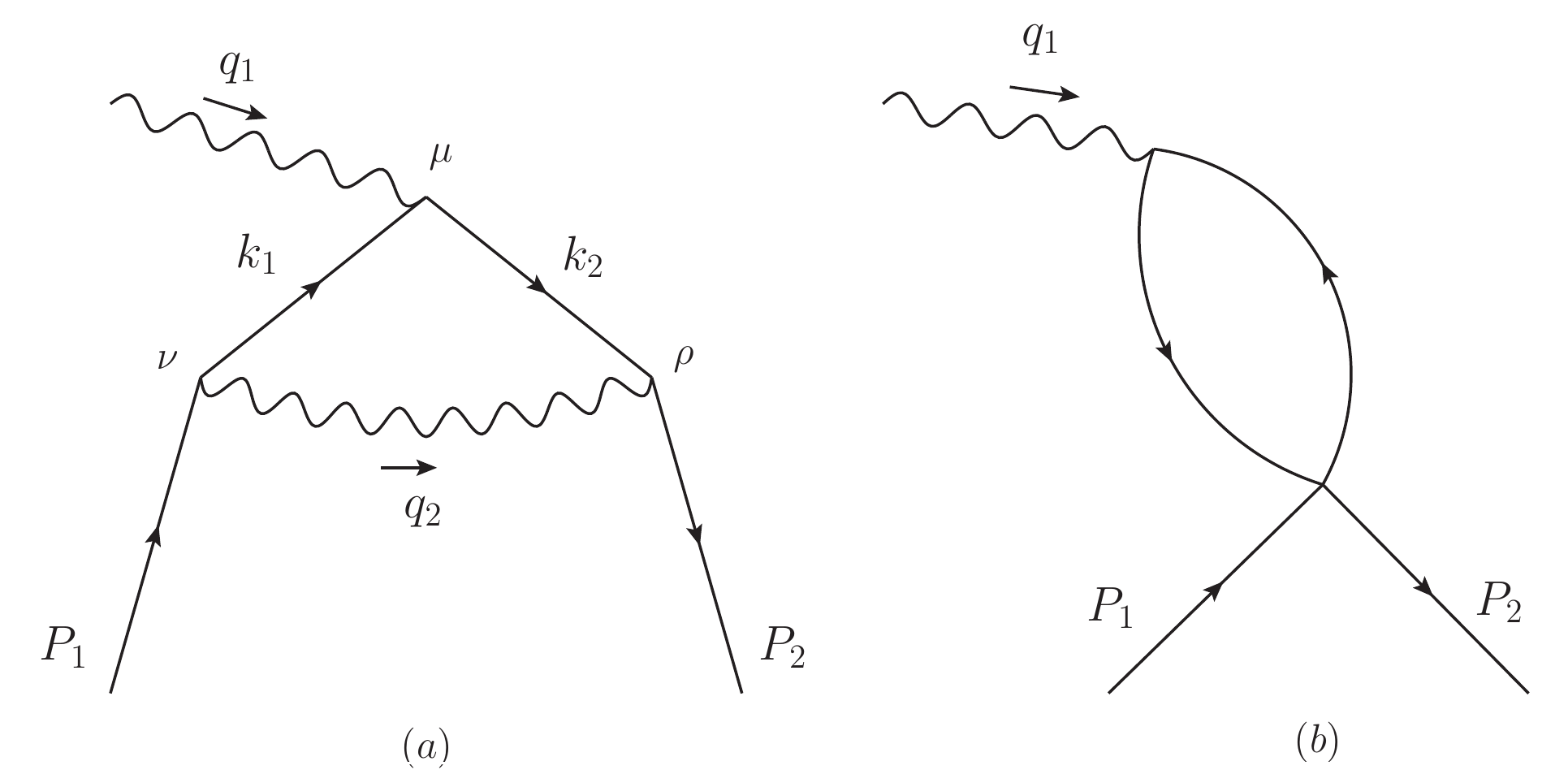}
\caption{Vertex correction at one-loop order in covariant QED: (a) triangle diagram; (b) swordfish diagram containing the vertex $V_I$.}
\label{trianglediagram}
\end{figure}

The one-loop vertex correction in the covariant approach with the light-cone-gauge condition $n_{\mu}A^{\mu}=0$
is described  in Fig.~\ref{trianglediagram}. Diagram (a) is the so-called triangle diagram 
due to the interaction term of the form $(V)^3$ in the interaction Hamiltonian \eqref{h-cov} and the diagram (b) is the so-called swordfish
diagram due to the term $V\,V_I$ containing an instantaneous four-fermion interaction. 

The  matrix element of the triangle diagram  reads
\[\Lambda=(2\pi)^4\delta^{(4)}\left(P_1+q_1-P_2\right){\left(-ie\right)}\varepsilon_{\mu}(q_1) J^{\mu}_{COV} \; , \]
with the  current $J^{\mu}$ given by
\begin{equation}
J^{\mu}_{COV}=\frac{e^2}{(2\pi)^4}\int d^4k_1\,\bar{u}\left(P_{2}\right)\gamma^{\rho} 
\mc{S}_F(k_2)
\gamma^{\mu}
\mc{S}_F(k_1)
\gamma^{\nu}\,
\mc{D}^{\nu\rho}_{T}(q_2)u(p_1)\; ,
\label{jmu}\end{equation}
where $\mc{S}_F(p)$  is the electron propagator, and $\mc{D}^{\nu\rho}_{T}(q)$ is the photon propagator in the light-cone gauge from Eq.~\eqref{proplc}.

\noindent
As discussed in Sec.~\ref{basics}, the third term in $\mc{D}^{\mu\nu}_{T}(q)$  generates a contribution which cancels out   that arising from  the $V_I$ interaction, corresponding to the diagram (b) in Fig.~\ref{trianglediagram}.

\noindent
As a result, 
the vertex correction in the covariant approach can effectively be  obtained by taking into account  the contribution from the triangle diagram in Fig.~\ref{trianglediagram} (a) alone, replacing $D_{T}^{\nu\rho}$ in \eqref{jmu} with the 
two-term  expression $D^{\nu\rho}$ of Eq.~\eqref{d2} for the photon propagator.
 
In the following, we show the equivalence of the covariant approaches  with light-front TOPT.
\\
In covariant theories, the momentum is conserved at each vertex; therefore we have
\begin{equation}
q_2= P_1-k_1 \; , \quad k_2=k_1+q_1 \; ,\quad P_2 = k_2+q_2 \; . \label{momcons}
\end{equation}
If we again apply the splitting of the momenta as in Eqs.~\eqref{splitting1}, \eqref{splitting2}, the fermion's and gauge boson's propagators get in turn decomposed into two parts, according to:
\begin{subequations}
\begin{align}
\mc{S}_{F}\left(k\right)=&\frac{i\left(k\!\!\!/+m\right)}{k^{2}-m^{2}+i\epsilon}=
\frac{i\left(\tilde{k}\!\!\!/+m\right)}{k^{2}-m^{2}+i\epsilon}+\frac{i\gamma^{+}}{2k^{+}}, \label{fermsplit} \\
\mc{D}^{\mu\nu}\left(q\right)=
&\frac{-i}{q^{2}+i\epsilon}\left(g^{\mu\nu}-\frac{q^{\mu}n^{\nu}+q^{\nu}n^{\mu}}{q^{+}}\right)\\
=&\frac{-i}{q^{2}+i\epsilon}\left(g^{\mu\nu}-\frac{\tilde{q}^{\mu}n^{\nu}+\tilde{q}^{\nu}n^{\mu}}{q^{+}}\right)
+\frac{in^{\mu}n^{\nu}}{\left(q^{+}\right)^{2}}=\frac{id^{\mu\nu}(\tilde{q})}{q^2+i\epsilon}+\frac{in^{\mu}n^{\nu}}{\left(q^{+}\right)^{2}}. \label{3term}
\end{align}
\end{subequations}
The first terms in both decompositions depend on the light-front energy components $k^{-}$ and $q^-$, and therefore they yield the propagating part; the second parts, instead, do not depend on the minus component of the momenta and hence they correspond to the non-propagating instantaneous particles \cite{Schoonderwoerd:1998qj}. It should be noticed that, to some extent, we recovered a three term propagator, but this time the numerator of the second term only depends on the on-shell component $\tilde{q}$ of the gauge-field momentum $q$.

As a consequence of the splitting in Eqs.~\eqref{fermsplit}-\eqref{3term}, the current $J^\mu_{COV}$ in Eq.~\eqref{jmu} can be rewritten as:
\begin{equation} 
J^{\mu}_{COV}=-\frac{ie^{2}}{\left(2\pi\right)^{4}}\int d^{4}k_1\,\bar{u}\left(P_{2}\right)\gamma^{\rho}\left(\frac{\tilde{\slashed{k}}_{2}+m}{k_{2}^{2}-m^{2}+i\epsilon}+\frac{\gamma^{+}}{2k_{2}^{+}}\right)\gamma^{\mu}\left(\frac{\tilde{\slashed{k}}_{1}+m}{k_{1}^{2}-m^{2}+i\epsilon}+\frac{\gamma^{+}}{2k_{1}^{+}}\right)\gamma^{\nu}\left(\frac{d_{\nu\rho}\left(\tilde{q}_{2}\right)}{q_{2}^{2}+i\epsilon}+\frac{n_{\nu}n_{\rho}}{(q_2^+)^2}\right)u\left(P_{1}\right) \; . \end{equation}
We can hence separate $J^{\mu}_{COV}$ into eight contributions, depending on different combinations of the propagating and instantaneous components of the propagators.
Out of these eight contributions, four are non-vanishing\footnote{We use the arguments in square brackets to specify which part (on-shell or off-shell) of the momenta enters in the numerator of the corresponding propagator.}:
\begin{subequations}
\begin{align}
& J_1^\mu\left[\tilde{k}_{1},\tilde{k}_{2},\tilde{q}_{2}\right]= -\frac{ie^{2}}{\left(2\pi\right)^{4}}\int d^{4}k_1\,\bar{u}\left(P_{2}\right)\gamma^{\rho}\left(\frac{\tilde{\slashed{k}}_{2}+m}{k_{2}^{2}-m^{2}+i\epsilon}\right)\gamma^{\mu}\left(\frac{\tilde{\slashed{k}}_{1}+m}{k_{1}^{2}-m^{2}+i\epsilon}\right)\gamma^{\nu}\frac{d_{\nu\rho}\left(\tilde{q}_{2}\right)}{q_{2}^{2}+i\epsilon}u\left(P_{1}\right) \; ,  \label{j1}\\
& J_2^\mu\left[\hat{k}_{1},\tilde{k}_{2},\tilde{q}_{2}\right]= -\frac{ie^{2}}{\left(2\pi\right)^{4}}\int d^{4}k_1\,\bar{u}\left(P_{2}\right)\gamma^{\rho}\left(\frac{\tilde{\slashed{k}}_{2}+m}{k_{2}^{2}-m^{2}+i\epsilon}\right)\gamma^{\mu}\frac{\gamma^{+}}{2k_{1}^{+}}\gamma^{\nu}\frac{d_{\nu\rho}\left(\tilde{q}_{2}\right)}{q_{2}^{2}+i\epsilon}u\left(P_{1}\right) \; ,  \\
& J_3^\mu\left[\tilde{k}_{1},\hat{k}_{2},\tilde{q}_{2}\right]= -\frac{ie^{2}}{\left(2\pi\right)^{4}}\int d^{4}k_1\,\bar{u}\left(P_{2}\right)\gamma^{\rho}\frac{\gamma^{+}}{2k_{2}^{+}}\gamma^{\mu}\left(\frac{\tilde{\slashed{k}}_{1}+m}{k_{1}^{2}-m^{2}+i\epsilon}\right)\gamma^{\nu}\frac{d_{\nu\rho}\left(\tilde{q}_{2}\right)}{q_{2}^{2}+i\epsilon}u\left(P_{1}\right) \; , \\
& J_4^\mu\left[\tilde{k}_{1},\tilde{k}_{2},\hat{q}_{2}\right]= -\frac{ie^{2}}{\left(2\pi\right)^{4}}\int d^{4}k_1\,\bar{u}\left(P_{2}\right)\gamma^{\rho}\left(\frac{\tilde{\slashed{k}}_{2}+m}{k_{2}^{2}-m^{2}+i\epsilon}\right)\gamma^{\mu}\left(\frac{\tilde{\slashed{k}}_{1}+m}{k_{1}^{2}-m^{2}+i\epsilon}\right)\gamma^{\nu}\frac{n_{\nu}n_{\rho}}{\left(q_{2}^{+}\right)^{2}}u\left(P_{1}\right) \; ,
\end{align}
\end{subequations}
while three are vanishing because of their Dirac matrix structures:
\begin{subequations}
\begin{align}
&J_5^\mu\left[\hat{k}_{1},\tilde{k}_{2},\hat{q}_{2}\right]  \propto\gamma^{\rho}\left(\tilde{\slashed{k}}_{2}+m\right)\gamma^{\mu}\gamma^{+}\gamma^{\nu}n_{\nu}n_{\rho}=0 \; , \\
&J_6^\mu\left[\tilde{k}_{1},\hat{k}_{2},\hat{q}_{2}\right]  \propto\gamma^{\rho}\gamma^{+}\gamma^{\mu}\left(\tilde{\slashed{k}}_{1}+m\right)\gamma^{\nu}n_{\nu}n_{\rho}=0 \; , \\
&J_7^\mu\left[\hat{k}_{1},\hat{k}_{2},\hat{q}_{2}\right]\propto  \gamma^{\rho}\gamma^{+}\gamma^{\mu}\gamma^{+}\gamma^{\nu}n_{\nu}n_{\rho}=0 \; .
\end{align}
\end{subequations}
There is one last term left, which is:
\begin{equation} J_8^\mu\left[\hat{k}_{1},\hat{k}_{2},\tilde{q}\right] \propto\gamma^{\rho}\gamma^+\gamma^{\mu}\gamma^{+}\gamma^{\nu}\left(g_{\nu\rho}-\frac{n_{\nu}\tilde{q}_{2_{\rho}}
+n_{\rho}\tilde{q}_{2_{\nu}}}{q_{2}^{+}}\right)=\gamma_{\nu}\gamma^+\gamma^-\gamma^+\gamma^\nu\delta^\mu_{-} \; . \label{j8}
\end{equation}
It is not vanishing by itself, but we notice that when we contract it with the polarization vector of the external photon, we obtain $\varepsilon^+(q) J_8^-$ which is zero in the light-cone gauge; therefore we can disregard this term, too.

The equivalence with the light-front TOPT approach can be established by integrating the covariant expression over the minus component of the momentum by residues.
We focus on the contribution $J_1^\mu$ in Eq.~\eqref{j1} as an example.
We can use again the momentum parametrization \eqref{symmnot}, which is this time valid for all the components of the momenta. The denominator of $J_{1}^{\mu}$ is then
\begin{equation}D_1\left[\tilde{k}_{1},\tilde{k}_{2},\tilde{q}_{2}\right]=\left[\left(k+\frac{\Delta}{2}\right)^{2}-m^{2}+i\epsilon\right]\left[\left(k-\frac{\Delta}{2}\right)^{2}-m^{2}+i\epsilon\right]\left[\left(P-k\right)^{2}+i\epsilon\right] \; . \label{den1}\end{equation}
If we also take advantage of the on-shell conditions for both the initial- and final-state electron, i.e. the first two identities in Eq.~\eqref{onshell},
the zeros of the denominator \eqref{den1} are:
\begin{equation}  \kappa_{\epsilon_{1}}^-=\kappa_{1}-\frac{i\epsilon}{2P^+(x-\xi)} \; , \quad  \kappa_{\epsilon_{2}}^-=\kappa_{2}-\frac{i\epsilon}{2P^+(x+\xi)}  \, , \quad \kappa_{\epsilon_{3}}^- = \kappa_{3}-\frac{i\epsilon}{2(x-1)P^+} \; , \label{poleseps}
\end{equation}
where $\kappa_{1}$, $\kappa_{2}$ and $\kappa_{3}$ are  defined in Eqs.~\eqref{poles1}-\eqref{poles2}.
Changing the variable of integration from $k_1\rightarrow k$ according to the relations \eqref{symmnot},
we see that the poles of the integrand in $J_1^{\mu}$ are 
distributed in the complex $k$ plane as shown in Table \ref{pole_dstrbt}.
\begin{table}
\begin{centering}
\begin{tabular}{|c|c|c|c|c|}
\hline
 & $x<-\xi$ & $-\xi<x<\xi$ & $\xi<x<1$ & $x>1$\tabularnewline
\hline
\hline
$\kappa_{\epsilon_{1}}^{-}$ & + & + & - & -\tabularnewline
\hline
$\kappa_{\epsilon_{2}}^{-}$ & + & - & - & -\tabularnewline
\hline
$\kappa_{\epsilon_{3}}^{-}$ & + & + & + & -\tabularnewline
\hline
\end{tabular}\protect\caption{Distribution of the poles in $J_{1}^{\mu}$ in the complex  plane: the symbols $+,-$ denote whether the pole
is located in the upper- or lower-half complex plane, respectively.}
\label{pole_dstrbt}
\par\end{centering}
\end{table}

The non-vanishing $k^-$-integral in $J_1^\mu$ therefore comes from the region $-\xi<x<1$: closing the circuit of integration in the upper- or lower-half complex plane, the integral is obtained from the residue in 
$\kappa_{\epsilon_{1}}^{-}$, with the result
\begin{align} J_1^\mu = & -\frac{e^{2}P^+}{(2\pi)^{3}}\int_{-\xi}^{\xi} dx\int d^2\pep{k}\,\bar{u}(P_2)\gamma^\rho\, \left(\frac{\tilde{\slashed{k}}_2+m}{\left(2P^+\right)^3\left(x^2-\xi^2\right)(x-1)}\right)\gamma^\mu\,\left(\frac{\tilde{\slashed{k}}_1+m}{\kappa_2^- -\kappa_1^-}\right)\gamma^\nu\,\left(\frac{d_{\nu\rho}(\tilde{q}_2)}{\kappa_{2}^- -\kappa_3^-}\right)u\left(P_{1}\right) \non \\
& +\frac{e^{2}P^+}{(2\pi)^{3}}\int_{\xi}^{1} dx\int d^2\pep{k}\,\bar{u}(P_2)\gamma^\rho\, \left(\frac{\tilde{\slashed{k}}_2+m}{\kappa_{3}^- -\kappa_2^-}\right)\gamma^\mu\,\left(\frac{\tilde{\slashed{k}}_1+m}{\kappa_3^- -\kappa_1^-}\right)\gamma^\nu\,\left(\frac{d_{\nu\rho}(\tilde{q}_2)}{\left(2P^+\right)^3\left(x^2-\xi^2\right)(x-1)}\right)u\left(P_{1}\right) \; . \non \\
\label{J1} \end{align}
We remark that the numerator remains unchanged after integration, since it does not depend on the minus component of $k$. 

The same procedure can be applied to terms $J_2^\mu$, $J_3^\mu$ and $J_4^\mu$. The integrand in $J_2^{\mu}$ has no poles in  $k_{1}^{-}$; therefore, after integration, it still results into the sum of two terms:
\begin{align} J_2^\mu = & -\frac{e^{2}P^+}{(2\pi)^{3}}\int_{-\xi}^{\xi} dx\int d^2\pep{k}\,\bar{u}(P_2)\gamma^\rho\left(\frac{\tilde{\slashed{k}}_2+m}{\left(2P^+\right)^3\left(x^2-\xi^2\right)(x-1)}\right)\gamma^\mu\,\gamma^+ \gamma^\nu\,\left(\frac{d_{\nu\rho}(\tilde{q}_2)}{\kappa_{2}^- -\kappa_3^-}\right)u\left(P_{1}\right) \non \\
& +\frac{e^{2}P^+}{(2\pi)^{3}}\int_{\xi}^{1} dx\int d^2\pep{k}\,\bar{u}(P_2)\gamma^\rho\left(\frac{\tilde{\slashed{k}}_2+m}{\kappa_{3}^- -\kappa_2^-}\right)\gamma^\mu\,\gamma^+ \gamma^\nu\,\left(\frac{d_{\nu\rho}(\tilde{q}_2)}{\left(2P^+\right)^3\left(x^2-\xi^2\right)(x-1)}\right)u\left(P_{1}\right) \non \\
& = \frac{e^{2}P^+}{(2\pi)^{3}}\int_{-\xi}^{1} dx\int d^2\pep{k}\,\bar{u}(P_2)\gamma^\rho\left(\frac{\tilde{\slashed{k}}_2+m}{\kappa_{3}^- -\kappa_2^-}\right)\gamma^\mu\,\gamma^+ \gamma^\nu\,\left(\frac{d_{\nu\rho}(\tilde{q}_2)}{\left(2P^+\right)^3\left(x^2-\xi^2\right)(x-1)}\right)u\left(P_{1}\right).
 \label{J2}\end{align}
The contributions $J_3^\mu$ and $J_4^\mu$, instead, are non-vanishing only in the regions $\xi<x<1$ and $-\xi<x<\xi$, respectively:
\begin{align} & J_3^\mu = \frac{e^{2}P^+}{(2\pi)^{3}}\int_{\xi}^{1} dx\int d^2\pep{k}\,\bar{u}(P_2)\gamma^\rho\,\gamma^+\gamma^\mu\left(\frac{\tilde{\slashed{k}}_1+m}{\kappa_{3}^- -\kappa_1^-}\right)\gamma^\nu\,\left(\frac{d_{\nu\rho}(\tilde{q}_2)}{\left(2P^+\right)^3\left(x^2-\xi^2\right)(x-1)}\right)u\left(P_{1}\right) \; , \label{J3}\\
& J_4^\mu = -\frac{e^{2}}{(2\pi)^{3}}\int_{-\xi}^{\xi} dx\int d^2\pep{k}\,\bar{u}(P_2)\gamma^\rho\, \left(\frac{\tilde{\slashed{k}}_2+m}{4\left(P^+\right)^3\left(x^2-\xi^2\right)(1-x)^2}\right)\gamma^\mu\left(\frac{\tilde{\slashed{k}}_1+m}{\kappa_2^- -\kappa_1^-}\right)\gamma^\nu\,n_{\nu}n_{\rho}u\left(P_{1}\right) \; . \label{J4} \end{align}

We are now ready to discuss the equivalence between the covariant approach and the light-front TOPT approach. Let us consider $J^{\mu}_{1}$ in Eq.~\eqref{J1}, which is the sum of two terms; if we compare them with the light-front TOPT results \eqref{ja} and \eqref{jb}, we see that these contributions coincide  with the sum of diagrams (a) and (b) in Fig.~\ref{lctopt}, where all particles in the intermediate states are propagating and we do not have instantaneous propagators.

The second contribution $J^{\mu}_{2}$ in covariant approach, Eq.~\eqref{J2}, coincides instead with Eq.~\eqref{jc}; notice that also in the covariant approach we can split this term into two, according to the value of the plus momentum flowing in the instantaneous propagator, thus reproducing the situation in the light-front TOPT case.

Finally, the current terms $J^{\mu}_{3}$ in \eqref{J3} and $J^{\mu}_{4}$ in  \eqref{J4} are exactly equivalent to their TOPT counterparts, namely Eq.~\eqref{jd} and Eq.~\eqref{je}, respectively; the second term, in particular, refers to the diagram with the instantaneously propagating photon. 

\noindent
The correspondence between the two approaches for the one-loop vertex correction is summarized in the Table \ref{J_eqvln}.

This result differs from the one found by Misra et al. in Ref.~\cite{Misra:2005dt}, where it is claimed that the three-term propagator is needed in order to obtain the equivalence. This is  due to the fact that in the calculation of the residue,
they do not evaluate $d^{\mu\nu}(q)$  at the pole position. By correctly calculating the residue, one would automatically include the contribution from the instantaneous photon, with no need of adding it separately in the third term.
Moreover, our results are more general because they apply to all the Lorentz components of the current, while the results in ~\cite{Misra:2005dt}
refer only to the contribution from the plus component.
\begin{table}
\centering{}%
\begin{tabular}{|c|c|}
\hline
Light-Front TOPT & Covariant Approach\tabularnewline
\hline
\hline
$J_{\left(a\right)}^{\mu}+J_{\left(b\right)}^{\mu}$ & $J_{1}^{\mu}$\tabularnewline
\hline
$J_{\left(c\right)}^{\mu}$ & $J_{2}^{\mu}$\tabularnewline
\hline
$J_{\left(d\right)}^{\mu}$ & $J_{3}^{\mu}$\tabularnewline
\hline
$J_{\left(e\right)}^{\mu}$ & $J_{4}^{\mu}$\tabularnewline
\hline
\end{tabular}\protect\caption{Correspondence between the different contributions to the triangle diagram 
in light-front TOPT and covariant approach.}\label{J_eqvln}
\end{table}

\section{One-loop self-energy diagrams}\label{selfen}
In this section we complete the proof of the equivalence between  light-front TOPT and covariant approaches in QED at one-loop order, by considering the self-energy diagrams for both the electron and the photon. We revisit the results in light-front TOPT which were also discussed in Refs.~\cite{Mustaki:1990im,Misra:2005dt}, and we prove the equivalence with the covariant approaches, fixing some 
imprecisions in the calculation of Ref.~\cite{Misra:2005dt}.

\subsection{Electron self-energy}

\begin{figure}[h]
\begin{center}
\includegraphics[scale=0.45]{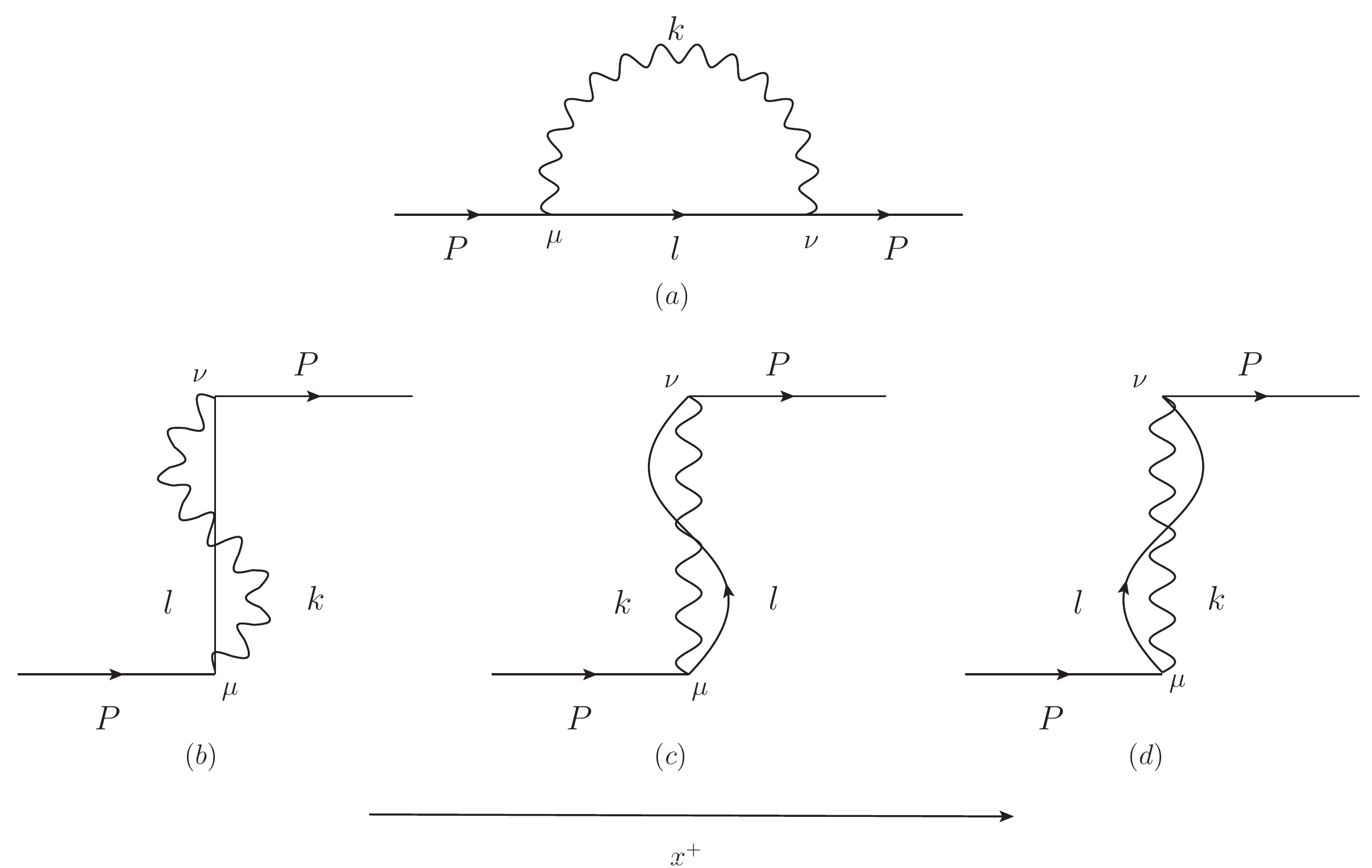}
\end{center}
\caption{
Diagrams for the electron self-energy in light-front TOPT at one-loop order, including the instantaneous exchange of fermions ((b)) and instantaneous exchange of photons ((c) and (d)). } 
\label{elself}
\end{figure}

If we consider the amplitude $T_{PP}$ of the transition matrix
\begin{equation}
T=V+V\frac{1}{P^- -H_0}V+\ldots \, , \qquad V= V_1+V_2+V_3 \label{T-matrix}
\end{equation}
between two electron states $\lvert P,s\rangle$, and $\lvert P,s'\rangle$, we can write
\begin{equation} T_{PP}\equiv\delta_{s,s'}\, \delta T=\bar{u}(P,s')\Sigma(P)u(P,s) \; , \end{equation} 
which defines the transition matrix $\Sigma (p)$\footnote{Our $\Sigma(P)$ has the dimension of a mass (consistently with, e.g., the Peskin-Schroeder notation~\cite{Peskin:1995ev} where $\Sigma(P)$ represents  the self-energy correction to the bare electron mass), at variance with the case of Ref.~\cite{Mustaki:1990im}, where $\Sigma(P)$ is dimensionless.}.
The $e^2$-order term of the perturbative expansion of $T_{PP}$ (and, consequently, of the  transition matrix) splits into three contributions, arising from the different terms of the interaction Hamiltonian:
\begin{equation}
\delta T\delta_{s,s'}=\left(\delta T_{a}+\delta T_{b}+\delta T_{c+d}\right)\delta_{s,s'}\equiv\bar{u}(P,s')\left[\Sigma^{(2)}_{a}(P)+\Sigma^{(2)}_{b}(P)+\Sigma^{(2)}_{c+d}(P)\right]u(P,s) \; . \end{equation}
The $\delta T_{a}$ term corresponds to the  contribution of second order in $V_{1}$, described by diagram (a) in Fig.~\ref{elself}. By using  $k^{+}=xP^{+}$  and $\left(l^+,\pep{l}\right)=\left((1-x)P^+,-\pep{k}\right)$, one finds 
\begin{equation} 
\delta T_{a}\delta_{s,s'}=\bar{u}_{s'}(P)\Sigma^{(2)}_{a}(P)u_{s}(P)=2e^2P^{+}\int\frac{d^{2}\pep{k}}{(4\pi)^3}\int_{0}^{1}dx\,\frac{1}{x(1-x)\left(P^{+}\right)^2}\bar{u}_{s'}(P)\frac{\gamma^{\nu}\left(\tilde{\slashed{l}}+m\right)\gamma^{\mu}d_{\mu\nu}(\tilde{k})}{P^{-}-k^{-}-l^{-}}u_{s}(P) \; . \label{ma}
 \end{equation}
The minus components can again be written from the on-shell conditions, i.e.
\begin{equation} P^-=\frac{m^2}{2P^{+}} \; , \quad k^-=\frac{\pep{k}^{2}}{2xP^+} \; , \quad l^-=\frac{\pep{k}^2+m^2}{2(1-x)P^+} \; .\label{onshell2}
\end{equation}

The $\delta T_{b}$ term, instead, refers to the contact interaction with an instantaneously propagating fermion, due to the contribution in $V_{1}V_{2}$ in light-front TOPT and
corresponds to diagram (b) in Fig.~\ref{elself}. It is given by
\begin{equation}
\delta T_{b}\delta_{s,s'}=\bar{u}_{s'}(P)\Sigma^{(2)}_{b}(P)u_{s}(P)=e^2\left(P^{+}\right)^2\int\frac{d^{2}\pep{k}}{(2\pi)^3}\int_{0}^{1}dx\,\frac{1}{x(1-x)\left(P^{+}\right)^2} \, \delta_{s,s'}\; . \label{mb}
\end{equation}
Finally, the third term $\delta T_{c+d}$ refers to the contact interaction with an instantaneously propagating photon, due to the contribution in $V_{1}V_{3}$ in light-front TOPT and corresponds to the sum of diagrams (c) and (d) in Fig.~\ref{elself}. 
It results into
\begin{equation}
\delta T_{c+d}\delta_{s,s'}=\bar{u}_{s'}(P)\Sigma^{(2)}_{c+d}(P)u_{s}(P)=e^2\left(P^{+}\right)^2\int\frac{d^{2}\pep{k}}{(2\pi)^3}\int_{0}^{\infty}dx\,\frac{1}{\left(P^+\right)^2}\left[\frac{1}{(1-x)^{2}}-\frac{1}{(1+x)^{2}}\right] \, \delta_{s,s'}\; . \label{mc}
\end{equation}

We now turn our attention to the covariant approach.
As we discussed in the case of the vertex correction, the calculation in light-front quantization can  effectively be performed   by taking into account only diagram (a) in Fig.~\ref{elself}, disregarding the contributions from instantaneous interactions, and using the two-term expression for the photon propagator. In this way, one has perfect equivalence with the formulation in the TOPT approach. 

The Feynman rules for diagram (a) in Fig.~\ref{elself} give
\begin{align}
-i\bar{u}_{s'}(P)\Sigma^{(2)}(P) \,u_{s}(P) &= -\frac{e^{2}}{\left(2\pi\right)^{4}}\int d^{4}k\,\bar{u}_{s'}(P)\gamma^{\nu}\frac{\slashed{l}+m}{l^{2}-m^{2}+i\epsilon}\gamma^{\mu}\frac{d_{\mu\nu}(k)}{k^{2}+i\epsilon} u_{s}(P) \non \\
&=-\frac{e^{2}}{\left(2\pi\right)^{4}}\int d^{4}k\,\bar{u}_{s'}(P)\gamma^{\nu}\left(\frac{\tilde{\slashed{l}}+m}{l^{2}-m^{2}+i\epsilon}+\frac{\gamma^{+}}{2l^{+}}\right)\gamma^{\mu}\left[\frac{d_{\mu\nu}\left(\tilde{k}\right)}{k^{2}+i\epsilon}-\frac{n_{\mu}n_{\nu}}{\left(k^{+}\right)^{2}}\right] u_{s} (P) \; ,
\end{align}
where we split the momenta according to Eqs. \eqref{splitting1} and \eqref{splitting2}; of course, we have $l=P-k$. One can rewrite $i\Sigma$ as the sum of the following four terms
\begin{subequations}
\begin{align}
-i\bar{u}_{s'}(P)\Sigma^{(2)}_{1}(P)u_{s}(P)\left[\tilde{l},\tilde{k}\right]= & -\frac{e^{2}}{\left(2\pi\right)^{4}}\int d^{4}k\,\bar{u}_{s'}\left(P\right)\gamma^{\nu}\frac{\tilde{l}\!\!\!/+m}{l^{2}-m^{2}+i\epsilon}\gamma^{\mu}\frac{d_{\mu\nu}\left(\tilde{k}\right)}{k^{2}+i\epsilon}u_{s}\left(P\right) \; , \label{sigma1}\\
-i\bar{u}_{s'}(P)\Sigma^{(2)}_{2}(P)u_{s}(P)\left[\hat{l},\tilde{k}\right]= & -\frac{e^{2}}{\left(2\pi\right)^{4}}\int d^{4}k\,\bar{u}_{s'}\left(P\right)\gamma^{\nu}\frac{\gamma^{+}}{2l^{+}}\gamma^{\mu}\frac{d_{\mu\nu}\left(\tilde{k}\right)}{k^{2}+i\epsilon}u_{s}\left(P\right) \; , \label{sigma2}\\
-i\bar{u}_{s'}(P)\Sigma^{(2)}_{3}(P)u_{s}(P)\left[\tilde{l},\hat{k}\right]= &\frac{e^{2}}{\left(2\pi\right)^{4}}\int d^{4}k\,\bar{u}_{s'}\left(P\right)\gamma^{\nu}\frac{\tilde{l}\!\!\!/+m}{l^{2}-m^{2}+i\epsilon}\gamma^{\mu}\frac{n_{\mu}n_{\nu}}{\left(k^{+}\right)^{2}}u_{s}\left(P\right) \; , \label{sigma3}\\
-i\bar{u}_{s'}(P)\Sigma^{(2)}_{4}(P)u_{s}(P)\left[\hat{l},\hat{k}\right]= & \frac{e^{2}}{\left(2\pi\right)^{4}}\int d^{4}k\,\bar{u}_{s'}\left(P\right)\gamma^{\nu}\gamma^{+}\gamma^{\mu}\frac{n_{\mu}n_{\nu}}{\left(k^{+}\right)^{2}}u_{s}\left(P\right) \; .\label{sigma4}
\end{align}
\end{subequations}
The contribution from $\Sigma_4$ in Eq.~\eqref{sigma4} is vanishing due to the structure of Dirac matrices.
For the remaining contributions, we proceed as outlined in the previous section by performing the in integration over $k^-$ by residues. The first term in Eq.~\eqref{sigma1} becomes

\begin{equation}
-i\bar{u}_{s'}(P)\Sigma^{(2)}_{1}(P)u_{s}(P)=-P^+\frac{ie^2}{2(2\pi)^3}\int_{0}^{1} dx\int d^2\pep{k} \, \bar{u}_{s'}(P)\frac{\gamma^{\nu}(\tilde{\slashed{k}}+m)\gamma^{\mu}d_{\mu\nu}(\tilde{q})}{P^{+}\left[\pep{k}^2+x^2 m^2\right]}u_{s}(P) \; .
\label{eselfcov}\end{equation}
It exactly coincides with Eq.~\eqref{ma}, via the conditions \eqref{onshell2}.

The terms in Eqs.~\eqref{sigma2} and ~\eqref{sigma3} are explicitly evaluated in Appendix \ref{app1}. Here we report only the final results given by
\begin{equation}
\bar{u}_{s'}(P)\Sigma^{(2)}_{2}(P)u_{s}(P)=\frac{e^2}{(2\pi)^3}\int d^2\pep{k}\int_{0}^{\infty}dx \,\frac{1}{x(1-x)}\delta_{s,s'} \; , \label{S2}
\end{equation}
\begin{equation}
\bar{u}_{s'}(P)\Sigma^{(2)}_{3}(P)u_{s}(P)=\frac{e^{2}}{(2\pi)^3}\int d^2\pep{k}\int_{0}^{\infty} dx\, \left[\frac{1}{(1-x)^2}-\frac{1}{(1+x)^2}\right]\delta_{s,s'} \; . \label{S3}
\end{equation}
By comparing these expressions with Eqs.~\eqref{mb} and \eqref{mc}, we can conclude that our calculation in the covariant approach perfectly reproduces the light-front TOPT results. 

We remark again that this result is in contrast with what is claimed in Ref.~\cite{Misra:2005dt}, where the equivalence with the TOPT result, however, is actually not achieved; this is because in Eq.~(58) of Ref.~\cite{Misra:2005dt} one should evaluate  $d_{\mu\nu}(k)$ at the pole position.

\subsection{Photon self energy}
\begin{figure}[h]
\centering{}\includegraphics[scale=0.45]{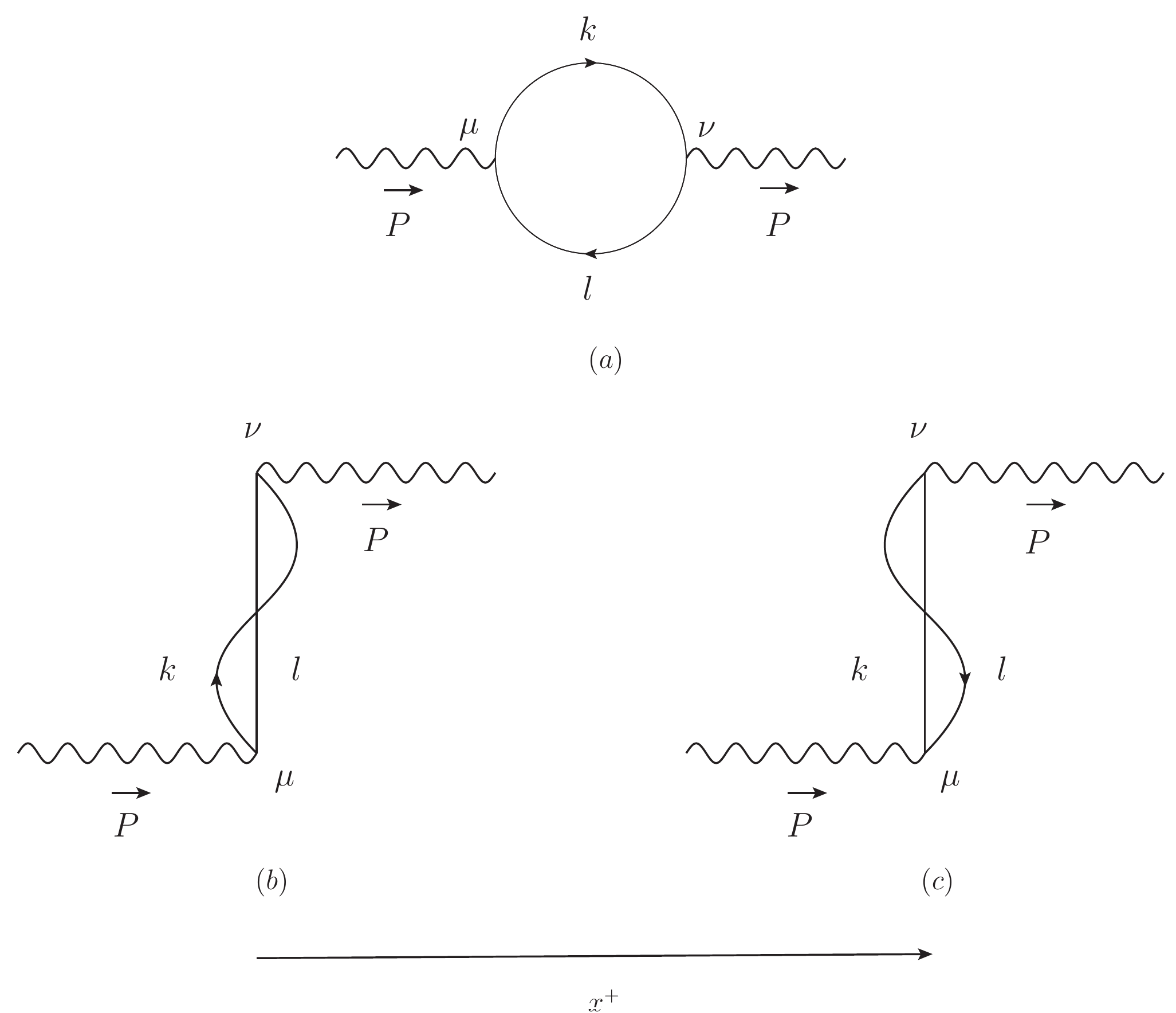}
\caption{
Diagrams for the photon self-energy in light-front TOPT at one-loop order, including the instantaneous exchange of electrons ((b)) and positrons ((c)). }
\label{phself}
\end{figure}

We finally discuss the photon self-energy corrections, corresponding to the diagrams in Fig.~\ref{phself} in light-front TOPT at order $e^2$.
Fllowing Ref.~\cite{Mustaki:1990im}, we denote with $T'_{PP}$  the amplitude at order $e^2$ of the transition matrix $T$ in Eq.~\eqref{T-matrix} between free-photon states with momentum and helicity $(P,\lambda)$ and $(P',\lambda')$, we can define  the self-energy correction to the fictitious photon mass as
\begin{equation} 
\delta\mu^2\delta_{\lambda,\lambda'}=2P^{+}T'_{PP}  \; ,
\end{equation}
and identify a tensor $\Pi^{\mu\nu}$ through the identity
\begin{equation} \delta\mu^2\delta_{\lambda,\lambda'}=\varepsilon_{\lambda',\mu}^{*}(P)\Pi^{\mu\nu}(P)\varepsilon_{\lambda,\nu}(P) \; . \end{equation}
It is important to notice that, in the above expressions, we need to consider only the physical degrees of freedom since both the incoming and the outgoing photons are real; therefore $\lambda,\lambda'=1,2$.
\\
We are then able to separate $\delta\mu^2$ (and consequently $\Pi^{\mu\nu}$) into two contributions. 
The first one arises from a  contribution in $(V_{1})^2$ in the  (light-front time-ordered) perturbative expansion, and corresponds to diagram (a) of Fig.~\ref{phself}: 
\begin{equation}
\delta\mu_{a}^2\delta_{\lambda,\lambda'}=\varepsilon_{\lambda',\mu}^{*}(P)\Pi_{a}^{\mu\nu}(P)\varepsilon_{\lambda,\nu}(P)=2e^2P^+\int \frac{d^{2}\pep{k}}{(4\pi)^3}\int_{0}^{1}dx\,\frac{1}{x(1-x)\left(P^+\right)^2}\frac{\mathrm{Tr}\left[\slashed{\varepsilon}^{*}_{\lambda'}(P)(\tilde{\slashed{k}}+m)\slashed{\varepsilon}_{\lambda}(P)(\tilde{\slashed{l}}-m)\right]}{P^- -k^- -l^-} \; . \label{mua}
\end{equation}
The on-shell conditions this time give
\begin{equation} 
P^- =0 \; , \quad k^-=\frac{\pep{k}^{2}+m^{2}}{2xP^+} \; , \quad l^-=\frac{\pep{k}^2+m^2}{2(1-x)P^+} \; .\label{onshell3}
\end{equation}

\noindent
The second 
contribution
 is due to the $V_{1}V_{2}$ interaction terms, corresponding to the sum of diagram (b) and (c) in Fig.~\ref{phself}, and turns out to be~\cite{Mustaki:1990im}
\begin{equation}
\delta\mu_{b+c}^2\delta_{\lambda,\lambda'}=\varepsilon_{\lambda',\mu}^{*}(P)\Pi_{b+c}^{\mu\nu}(P)\varepsilon_{\lambda,\nu}(P)=e^2\int\frac{d^{2}\pep{k}}{(2\pi)^3}\int_{0}^{\infty} dx\,\left[\frac{1}{1-x}-\frac{1}{1 +x}\right]\delta_{\lambda,\lambda'} \; . \label{mubc}
\end{equation}

In the covariant approach, we proceed as discussed in sects.~\ref{triangle} and \ref{selfen}, and we apply the Feynman rules for the calculation of  the diagram (a) in Fig.~\ref{phself}:
\begin{equation}
i\varepsilon^{*}_{\lambda',\nu}(P)\Pi^{\mu\nu}(P)\varepsilon_{\lambda,\mu}(P)= \frac{e^{2}}{\left(2\pi\right)^{4}}\int d^{4}k\,\varepsilon_{\lambda',\nu}^{*}(P)\frac{\mathrm{Tr}\left[\gamma^{\nu}(-\slashed{l}+m)\gamma^{\mu}(\slashed{k}+m)\right]}{\left(l^{2}-m^{2}+i\epsilon\right)\left(k^{2}-m^{2}+i\epsilon\right)}\,\varepsilon_{\lambda,\mu}(P) \; .\end{equation}
Using the decompositions for the momenta in Eqs.~\eqref{splitting1} and \eqref{splitting2}, we can identify the following four contributions to the self-energy:

\begin{align}
i\varepsilon^{*}_{\lambda',\nu}(P)\Pi^{\mu\nu}_{1}(P)\varepsilon_{\lambda,\mu}(P)\left[\tilde{k},\tilde{l}\right]= & -\frac{e^{2}}{\left(2\pi\right)^{4}}\int d^{4}k_{1}\,\varepsilon^{*}_{\lambda',\nu}(P)\frac{\mathrm{Tr}\left[\gamma^{\nu}\left(\tilde{\slashed{l}}-m\right)\gamma^{\mu}\left(\tilde{\slashed{k}}+m\right)\right]}{\left(l^{2}-m^{2}+i\epsilon\right)\left(k^{2}-m^{2}+i\epsilon\right)}\,\varepsilon_{\lambda,\mu}(P)\;  \non\\
= & -\frac{ie^2}{2(2\pi)^3}\int_{0}^{1}dx\int d^{2}\pep{k}\, \frac{\mathrm{Tr}\left[\slashed{\varepsilon}_{\lambda}(P) \left(\tilde{\slashed{k}}+m\right)\slashed{\varepsilon}^{*}_{\lambda'}(P)\left(\tilde{\slashed{l}}-m\right)\right]}{\pep{k}^2+m^2} \; , \label{pi1} \end{align}
\begin{subequations}
\begin{align}
i\varepsilon^{*}_{\lambda',\nu}(P)\Pi^{\mu\nu}_2(P)\varepsilon_{\lambda,\mu}(P)\left[\hat{k},\tilde{l}\right]= & -\frac{e^{2}}{\left(2\pi\right)^{4}}\int d^{4}k\,\varepsilon^{*}_{\lambda',\nu}(P)\frac{\mathrm{Tr}\left[\gamma^{\nu}\left(\tilde{\slashed{l}}-m\right)\gamma^{\mu}\gamma^{+}\right]} {2k^{+}\left(l^{2}-m^{2}+i\epsilon\right)}\varepsilon_{\lambda,\mu}(P) \;  \label{pi2a} \\
= & -\frac{ie^{2}}{2\left(2\pi\right)^{3}}\int d^2\pep{k}\int dx\,\frac{1}{x}\text{sgn}\left(1-x\right)\delta_{\lambda,\lambda'} \; ,\label{pi2}
\end{align}
\end{subequations}
\begin{subequations}
\begin{align}
i\varepsilon^{*}_{\lambda',\nu}(P)\Pi^{\mu\nu}_3(P)\varepsilon_{\lambda,\mu}(P)\left[\tilde{k},\hat{l}\right]= & \frac{e^{2}}{\left(2\pi\right)^{4}}\int d^{4}k\,\varepsilon^{*}_{\lambda',\nu}(P)\frac{\mathrm{Tr}\left[\gamma^{\nu}\gamma^{+}\gamma^{\mu}\left(\tilde{\slashed{k}}+m\right)\right]}{2l^{+}\left(k^{2}-m^{2}+i\epsilon\right)}\varepsilon_{\lambda,\mu}(P) \;  \label{pi3a} \\
= &\frac{ie^{2}}{2\left(2\pi\right)^{3}}\int d^2\pep{k}\int dx\,\frac{1}{1-x}\text{sgn}\left(x\right)\delta_{\lambda,\lambda'}\label{pi3} \; , 
\end{align}
\end{subequations}
\begin{equation}i\varepsilon^{*}_{\lambda',\nu}(P)\Pi^{\mu\nu}_4(P)\varepsilon_{\lambda,\mu}(P)\left[\hat{k},\hat{l}\right]= \frac{e^{2}}{\left(2\pi\right)^{4}}\int d^{4}k\,\frac{1}{2l^{+}}\frac{1}{2k^{+}}\varepsilon^{*}_{\lambda',\nu}(P)\mathrm{Tr}\left[\gamma^{\nu}\gamma^{+}\gamma^{\mu}\gamma^{+}\right]\varepsilon_{\lambda,\mu}(P)  \; . \label{pi4}
\end{equation}

The final results for $\Pi^{\mu\nu}_1$ in Eq.~\eqref{pi1} is obtained following the same procedure adopted for the electron self-energy with the on-shell condition $P^2=0$; the derivation of Eqs.~\eqref{pi2} and ~\eqref{pi3} can be found in Appendix \ref{app2}. 

\noindent
Finally, the contribution from $\Pi^{\mu\nu}_4$ in \eqref{pi4} is vanishing. 
This follows from 
\begin{equation} 
\mathrm{Tr}\left[\gamma^{\nu}\gamma^{+}\gamma^{\nu}\gamma^{+}\right]=2g^{\nu+}g^{\mu+} \label{trace}\end{equation}
and the contraction with the polarization vectors, since
in the light-cone gauge  $\varepsilon^{+}(P)=0$.
If we use the substitution $x\rightarrow 1-x'$ in Eq.~\eqref{pi2} and then sum it with Eq.~\eqref{pi3}, it is straightforward to check that
\begin{align} i\varepsilon^{*}_{\lambda',\nu}(P)\left[\Pi^{\mu\nu}_{2}(P)+\Pi^{\mu\nu}_{3}\right]\varepsilon_{\lambda,\mu}(P)&=\frac{e^{2}}{(2\pi)^3}\int d^{2}\pep{k}\left[\int_{0}^{\infty} dx\, \frac{1}{1-x} -\int_{-\infty}^{0} dx\,\frac{1}{1-x}\right]\non \\
&=\frac{e^{2}}{(2\pi)^3}\int d^{2}\pep{k}\int_{0}^{\infty}dx\,\left(\frac{1}{1-x}-\frac{1}{1+x}\right) \; . \label{pi23}\end{align}

By comparing Eqs.~\eqref{pi1} and \eqref{pi23} with the TOPT results \eqref{mua}-\eqref{mubc}, we once again find perfect agreement between the two approaches. This argument concludes the proof of the equivalence at one-loop level.

\section{Conclusions}\label{conclusions}

The present work clarifies some of the subtleties concerning the use of the
non-covariant light-cone gauge in the formalism of covariant
theories, with particular regard to the form of the gauge-field propagator. The
comparison between light-front TOPT and covariant formulation of QED at
one-loop level turns out to be a suitable proving ground to address these
questions. \\ 

We re-examined the derivation of the vertex correction, fermion self-energy
and vacuum polarization at one-loop level in QED, systematically applying the
splitting of the propagators into their on-shell and off-shell components. We
applied the standard technique of integration by residues to show how the
covariant expressions in light-cone gauge reduce, after integration over the
light-cone energy in the loop, to contributions from differently time-ordered
diagrams. 

We assumed as a starting point for the calculation in the covariant formalism
the three-term photon propagator of Eq.~\eqref{threeterm}; this is equivalent to considering nly the transverse degrees of freedom as the propagating ones for the free photon. However, the use of a
non-covariant gauge modifies the interaction Hamiltonian and it turns out that the
third term of the propagator is exactly canceled by an instantaneous
interaction term in the new Hamiltonian. This is a general feature of any
axial gauge, as shown in Ref.~\cite{Ji:2014vha}; it becomes therefore natural
to consider only the two-term propagator \eqref{twoterm} as a starting point
also for the covariant approach, omitting at the same time the diagrams with
instantaneous interactions. \\

The time-ordered diagrams containing instantaneously-propagating photons,
which are an intrinsic property of the light-front theory, are recovered in
the covariant approach thanks to the off-shell component of the photon
propagator. The on-shell part, instead, matches the contribution of
forward-propagating particles in time-ordered diagram, consistently with the
fact that in TOPT all particles must be taken on shell. \\

We believe that our consistent treatment of the two approaches generalizes the
previous works in the literature concerning this topic
\cite{Weinberg:1966jm,Schoonderwoerd:1998qj,Schoonderwoerd:1998iy,Ji:2014vha,Misra:2005dt,Patel:2010vd}. However,
our results differ from the findings of Refs.~\cite{Misra:2005dt,Ji:2014vha},
where it was claimed instead that one needs to start with a three-term gauge-field
propagator in the the covariant formula in order to generate the diagrams with
instantaneous photons. We proved instead that the contribution for the
instantaneously propagating particle is already contained in the off-shell
part of the second term. \\ 

We mention that a more general extension of this work could be achieved by
interpolating the light-front and instant-form coordinates, as done in
\cite{Ji:2014vha} for the scalar QED case, and working in a generic axial
gauge $n_{\mu}A^{\mu}=0$. 

\begin{acknowledgements}

This work was partially supported by the European Research Council (ERC) under
the  European Union's 
Horizon 2020 research and innovation programme (grant agreement No. 647981,
3DSPIN)

\end{acknowledgements}

\appendix

\section{}\label{app1}
In this Appendix we  explicitly derive the results for the contributions to the electron self-energy from
 $\Sigma_2$ and $\Sigma_3$ in Eqs.~\eqref{S2} and \eqref{S3}, respectively. 
 
We start from the contribution $\Sigma_2$. 
 The matrix element in the numerator of the integrand in Eq.~\eqref{sigma2} can be rewritten as
\begin{equation} 
\bar{u}_{s'}(P)\gamma^{\nu}\gamma^{+}\gamma^{\mu}d_{\mu\nu}(\tilde{k})u_{s}(P)=2\bar{u}_{s'}(P)\gamma^{+}u_{s}(P)=4P^{+}\delta_{s,s'}, 
\end{equation}
where we used the relations
\[\gamma_{\mu}\gamma^{\nu}\gamma^{\mu}=-2\gamma^{\nu} \; , \quad \bar{u}_{s'}(P)\gamma^{+}u_{s}(P)=2P^{+}\delta_{s,s'} \; . \]
As a result we can write
\begin{equation} -i\bar{u}_{s'}(P)\Sigma_{2}(P)u_{s}(P)=-\frac{e^{2}P^{+}}{2m(2\pi)^{3}}\delta_{s,s'}\int d^{2}\pep{k}\int dx\, \frac{4P^{+}}{2(1-x)P^{+}} I_{1} \; , \label{S2a}\end{equation}
where in $I_1$ we have isolated the integral over $k^-$, i.e.
\[ I_{1}=\int \frac{dk^-}{2\pi}\frac{1}{k^{2}+i\epsilon} \; . \]  
By introducing the new variable
\begin{equation} u=\frac{1}{k^{-}} \; , \label{u}\end{equation}
we can rewrite $I_1$ as
\begin{equation}
 I_1=-\frac{1}{2xP^+}\int \frac{du}{2\pi}\frac{1}{u}\frac{1}{\left[1-u\left(\frac{\pep{k}^{2}-i\epsilon}{2xP^{+}}\right)\right]} \; , 
 \label{i1}
 \end{equation}
which shows poles for $u=0$ (i.e. $k^{-}=\infty$) and $u=\frac{2xP^{+}}{\pep{k}^2-i\epsilon}\equiv u_x$. 
We regularize the first singularity by substituting
\begin{equation} \frac{1}{u}\rightarrow \frac{1}{2}\left(\frac{1}{u+i\delta}+\frac{1}{u-i\delta}\right) \; . \label{ureg}\end{equation}
As a result, we have
\begin{equation} I_1=-\frac{1}{4xP^+}\int \frac{du}{2\pi}\frac{1}{\left[1-u\left(\frac{\pep{k}^{2}-i\epsilon}{2xP^{+}}\right)\right]}\left(\frac{1}{u+i\delta}+\frac{1}{u-i\delta}\right) \; . \end{equation}
When considering the first term in round brackets, the first singularity now falls into the lower half-plane (at $u=-i\delta$); therefore we will obtain a nonzero result when $u_x$ is in the upper half-plane, which means for $x>0$. The specular situation holds for the second term in round brackets, which will then contribute only for $x<0$. 
\\
If we choose the contour of  integration for the first and second term enclosing the $-i\delta$  and $+i\delta$ poles, respectively,  and then take the limit $\delta\rightarrow 0$, it is easy to see that we come up with the following result

\begin{equation} I_1=\frac{i}{4xP^{+}}\left[\theta(x)-\theta(-x)\right]=\frac{i}{4xP^{+}} \mathrm{sgn}(x) \; . \label{I1}\end{equation}
As we insert Eq.~\eqref{I1} into \eqref{S2a}, we finally get
\begin{equation}
-i\bar{u}_{s'}(P)\Sigma_{2}(P)u_{s}(P)=-\frac{ie^2}{4m(2\pi)^3}\delta_{s,s'}\int d^{2}\pep{k}\int dx \,\frac{1}{x(1-x)}\mathrm{sgn}(x)=-\frac{ie^2}{2m(2\pi)^3}\int d^2\pep{k}\int_{0}^{\infty}dx \,\frac{1}{x(1-x)}\delta_{s,s'} \; .
\end{equation}

For the calculation of  the contribution $\Sigma_3$, we start
 from Eq.~\eqref{sigma3} and rewrite  the matrix element in the numerator 
 of the integrand as
\begin{equation}
\bar{u}_{s'}(P)\gamma^{+}(\tilde{\slashed{l}}+m)\gamma^{+}u_{s}(P)=l^{+}\bar{u}_{s'}(P)\gamma^{+}\gamma^{-}\gamma^{+}u_{s}(P)=2(1-x)P^{+}\bar{u}_{s'}(P)\gamma^{+}u_{s}(P)=4(1-x)\left(P^+\right)^2 \delta_{s,s'} \; .\end{equation}
As a result, we have
\begin{equation}
-i\bar{u}_{s'}(P)\Sigma_{3}(P)u_{s}(P)=\frac{e^{2}P^{+}}{2m(2\pi)^{3}}\delta_{s,s'}\int d^{2}\pep{k}\int dx \,\frac{4(1-x)\left(P^{+}\right)^2}{\left(xP^{+}\right)^2} I_2 \; ,\label{S3a}
\end{equation}
where  the integral $I_2$ over $k^{-}$ is given by
\begin{equation} I_2=\int \frac{dk^-}{2\pi}\frac{1}{(P-k)^{2}-m^{2}+i\epsilon} \; .\label{I2}\end{equation}
Changing the  variable of integration as in \eqref{u}, and using the on-shell condition $P^2=m^2$, one obtains
\begin{equation}
I_2=-\frac{1}{2(1-x)P^+}\int\frac{du}{2\pi}\frac{1}{u}\frac{1}{\left[1-u\,\frac{xm^{2}+\pep{k}^2-i\epsilon}{2(x-1)P^+}\right]} \; .
\end{equation}
We can henceforth proceed as before, replacing $1/u$ according to \eqref{ureg};
the integration around the pole at $u=-i\delta$ will now give a nonzero result only when $x-1>0$, while the integration around the pole at $u=i\delta$ contributes when $x-1<0$. By taking the limit $\delta\rightarrow 0$, one finds
\begin{equation}
I_2=\frac{i}{4(x-1)P^{+}}\left[\theta(x-1)-\theta(1-x)\right]=-\frac{i}{4(x-1)P^{+}}\mathrm{sgn}(1-x) \; . \label{I2b}
\end{equation}
Inserting this back in \eqref{S3a}, one comes up with
\begin{equation}
-i\bar{u}_{s'}(P)\Sigma_{3}(P)u_{s}(P)=\frac{ie^2}{2m\left(2\pi\right)^3}\delta_{s,s'}\int d^2\pep{k}\int dx\,\frac{1}{x^2}\,\mathrm{sgn}(1-x) \; . \label{S3b}
\end{equation}
Note that, since $1/x^{2}$ is an even function of $x$, we have:
\begin{align} \int dx\, \frac{1}{x^{2}}\,\mathrm{sgn}(1-x)&=-\int_{-\infty}^{1}dx\, \frac{1}{x^{2}}+\int_{1}^{\infty}dx\, \frac{1}{x^{2}}\non \\
&=-\int_{-1}^{\infty}dx\, \frac{1}{x^{2}}+\int_{1}^{\infty}dx\, \frac{1}{x^{2}}=-\int_{0}^{\infty}dx\,\frac{1}{(1-x)^2}+\int_{0}^{\infty}dx\,\frac{1}{(1+x)^{2}} \; .\end{align}
We therefore conclude that Eq.~\eqref{S3b} becomes
\begin{equation} \bar{u}_{s'}(P)\Sigma_{3}(P)u_{s}(P)=\frac{e^{2}}{2m(2\pi)^3}\delta_{s,s'}\int d^2\pep{k}\int_{0}^{\infty} dx\, \left[\frac{1}{(1-x)^2}-\frac{1}{(1+x)^2}\right] \; . 
\end{equation}

\section{}\label{app2}
We finally prove the derivation of Eqs.~\eqref{pi2} and \eqref{pi3} from \eqref{pi2a} and \eqref{pi3a}, respectively. Notice that (thanks to the invariance of the trace under cyclic permutations), Eqs.~\eqref{pi2a} and \eqref{pi3a} can be obtained one from the other via the exchange $k\leftrightarrow l$ and putting an extra minus sign in front, so we need to evaluate only one of the two equations. We hence start by considering the numerator of \eqref{pi2a}
\begin{equation}
\mathrm{Tr}\left[\gamma^{\nu}\tilde{\slashed{l}}\gamma^{\mu}\gamma^+\right]=l^{+}\mathrm{Tr}\left[\gamma^{\nu}\gamma^-\gamma^{\mu}\gamma^+\right]+l^{-}\mathrm{Tr}\left[\gamma^{\nu}\gamma^+\gamma^{\mu}\gamma^+\right]-l_{i}\mathrm{Tr}\left[\gamma^{\nu}\gamma^{i}\gamma^{\mu}\gamma^+\right] \; . \end{equation}
Out of these three terms, the second one can be neglected thanks to the same argument we made for \eqref{pi4}, while the third one can be dropped since $l_{i}=-k_{i}$ ($i=1,2$) and we are integrating over $d^2\pep{k}$. Hence in our numerator we are left with
\begin{equation} \varepsilon_{\lambda',\,\nu}^{*}(P)\mathrm{Tr}\left[\gamma^{\nu}\gamma^-\gamma^{\mu}\gamma^{+}\right]\varepsilon_{\lambda,\mu}(P)=4\delta_{\lambda,\lambda'} \; , \end{equation}
where we used 
\[\varepsilon_{\lambda',\,\mu}^{*}(P)\varepsilon_{\lambda}^{\mu}(P)=g_{\lambda,\lambda'}=-\delta_{\lambda,\lambda'} \; , \qquad \text{for} \; \; \lambda,\lambda'=1,2 \; . \]
Therefore we can rewrite
\begin{equation}
i\varepsilon^{*}_{\lambda',\nu}(P)\Pi^{\mu\nu}_2(P)\varepsilon_{\lambda,\mu}(P)=-\frac{4e^{2}P^{+}}{(2\pi)^{3}}\int d^{2}\pep{k}\int dx\,\frac{(1-x)P^+}{xP^+}\delta_{\lambda,\lambda'}I_{2}\; , \label{pi2c}\end{equation}
where the integral $I_{2}$ coincides with \eqref{I2} and yields the result \eqref{I2b}; putting the latter into \eqref{pi2c}, we get Eq.~\eqref{pi2}. The substitution $k\rightarrow l$ now reduces to the exchange $x\rightarrow 1-x$; putting also a minus sign on top, we recover Eq.~\eqref{pi3}.

\clearpage
\bibliographystyle{myrevtex}
\bibliography{Equiv_biblio}

\end{document}